\documentstyle[psfig]{mn2e}
\newcommand{\etal}{{et al.~}}

\newcommand{\gta}{\ga}

\newcommand{\kms}{\>{\rm km}\,{\rm s}^{-1}}

\newcommand{\Mpc}{\>{\rm Mpc}}

\newcommand{\msun}{\>{\rm M_{\odot}}}
\newcommand{\Lsun}{\>{\rm L_{\odot}}}

\newcommand{\beq}{\begin{equation}}
\newcommand{\eeq}{\end{equation}}

\newcommand{\mpch}{\>h^{-1}{\rm {Mpc}}}

\newcommand{\apj}{ApJ}

\newcommand{\aj}{AJ}
\newcommand{\mnras}{MNRAS}

\newdimen\hssize
\newdimen\hdsize 
\begin{document}
            

\title[The Cross-Correlation between Galaxies and Groups]
      {The Cross-Correlation between Galaxies and Groups:
       Probing the Galaxy Distribution in and around Dark Matter Haloes}    
\author[X. Yang et al.]
       {Xiaohu Yang$^{1}$ \thanks{E-mail: xhyang@astro.umass.edu}, 
       H.J. Mo$^{1}$, Frank C. van den Bosch$^{2}$, Simone M. Weinmann$^{2}$, 
\newauthor Cheng Li$^3$, Y.P. Jing$^{4}$\\
      $^1$Department of Astronomy, University of Massachusetts,
           Amherst MA 01003-9305, USA\\
      $^2$Department of Physics, Swiss Federal Institute of
           Technology, ETH H\"onggerberg, CH-8093, Zurich, Switzerland\\ 
      $^3$Center for Astrophysics, University of Science and
           Technology of China, Hefei, 230026, China\\
      $^4$Shanghai Astronomical Observatory; the Partner Group of MPA,
           Nandan Road 80, Shanghai 200030, China}


\date{}


\maketitle

\label{firstpage}


\begin{abstract}
  We  determine the  cross-correlation function  between  galaxies and
  galaxy  groups,  using both  the  Two-Degree  Field Galaxy  Redshift
  Survey (2dFGRS) and the Sloan Digital Sky Survey (SDSS).  Groups are
  identified using  the halo-based group  finder developed by  Yang et
  al., which is optimized to  associate those galaxies to a group that
  belong  to  the  same  dark  matter halo.   Our  galaxy-group  cross
  correlation function  is therefore  a surrogate for  the galaxy-halo
  cross  correlation function.   We study  the cross-correlation  as a
  function of group mass, and as a function of the luminosity, stellar
  mass, colour, spectral type and  specific star formation rate of the
  galaxies.   All  these  cross-correlation  functions  show  a  clear
  transition  from the  `1-halo' to  the `2-halo'  regimes on  a scale
  comparable to the  virial radius of the groups  in consideration. On
  scales  larger   than  the  virial   radius,  all  cross-correlation
  functions  are roughly  parallel,  consistent with  the linear  bias
  model.  In  particular, the  large scale correlation  amplitudes are
  higher  for  more  massive  groups,  and  for  brighter  and  redder
  galaxies.   In the `1-halo'  regime, the  cross-correlation function
  depends strongly on the definition  of the group center. We consider
  both a luminosity-weighted center (LWC)  and a center defined by the
  location  of  the brightest  group  galaxy  (BGC).   With the  first
  definition,  the bright  early-type galaxies  in massive  groups are
  found to be more  centrally concentrated than the fainter, late-type
  galaxies. Using the BGC, and excluding the brightest galaxy from the
  cross correlation analysis, we  only find significant segregation in
  massive  groups  ($M   \gta  10^{13}h^{-1}\msun$)  for  galaxies  of
  different  spectral types  (or  colours or  specific star  formation
  rates). In  haloes with masses $\la 10^{13}h^{-1}\msun$,  there is a
  significant  deficit of  bright satellite  galaxies.   Comparing the
  results  from the  2dFGRS with  those obtained  from  realistic mock
  samples, we find that the distribution of galaxies in groups is much
  less concentrated  than dark matter haloes predicted  by the current
  $\Lambda$CDM model.
\end{abstract}




\section{Introduction}

In  the  standard  cold  dark matter  (CDM)  cosmogony,  gravitational
instability  of the  cosmic density  field leads  to the  formation of
virialized  clumps of  dark  matter, called  dark  matter haloes,  and
galaxies are assumed  to form in these haloes  through gas cooling and
condensation.   One  of the  ultimate  challenges  in astrophysics  is
therefore  to obtain  a detailed  understanding of  how  galaxies with
different physical  properties occupy dark matter  haloes of different
mass.   This galaxy/dark  halo  connection is  an  imprint of  various
complicated  physical  processes  governing  galaxy formation,  and  a
detailed quantification of this connection is an important key towards
understanding galaxy formation and evolution within the CDM cosmogony.

To  quantify  the  relationship  between  haloes  and  galaxies  in  a
statistical  way,  one  can  specify  the  so-called  halo  occupation
distribution, $P(N \vert M)$, which  gives the probability to find $N$
galaxies (with some specified properties) in a halo of mass $M$.  This
occupation  distribution   can  be  constrained  using   data  on  the
clustering  properties of  galaxies,  as it  completely specifies  the
galaxy  bias on  large  scales.  In  the  last couple  of years,  this
approach  has  been  used   extensively  to  study  galaxy  occupation
statistics  and large  scale  structure (Jing,  Mo  \& B\"orner  1998;
Peacock  \& Smith  2000; Seljak  2000; Scoccimarro  \etal  2001; Jing,
B\"orner \& Suto 2002; Berlind  \& Weinberg 2002; Bullock, Wechsler \&
Somerville 2002;  Scranton 2002; Kang  \etal 2002; Marinoni  \& Hudson
2002; Zheng \etal 2002;  Magliocchetti \& Porciani 2003; Berlind \etal
2003; Yang \etal 2004; Zehavi  \etal 2004a,b; Zheng \etal 2004).  In a
series of papers, Yang, Mo \&  van den Bosch (2003) and van den Bosch,
Yang  \&   Mo  (2003)  extended  this  halo   occupation  approach  by
introducing the conditional luminosity  function (CLF), which allows a
study  of  the  halo  occupation  statistics  as  function  of  galaxy
luminosity and type.  So far,  the CLF formalism has provided a wealth
of  information  regarding  the  galaxy-dark matter  connection.   For
example, Yang \etal  (2003) and van den Bosch  \etal (2003) found that
the halo mass-to-galaxy light  ratio is a strongly non-linear function
of halo mass, indicating that the star formation efficiency depends on
halo mass  in a  complicated way.  Mo  \etal (2004)  made predictions,
based  on the  CLF, for  the  environmental dependence  of the  galaxy
luminosity function. Croton \etal (2005) showed that these predictions
are in excellent agreement with the observational data. This indicates
that there is no environment dependence beyond the halo virial radius.
Yang \etal (2005c), using galaxy groups identified from the Two-Degree
Field Galaxy Redshift Survey (2dFGRS, Colless \etal 2001; 2003) with 
the halo-based group finder developed in Yang \etal (2005a), found that
the halo occupation statistics derived directly from the galaxy groups
are  perfectly  consistent  with  those  obtained from  the  CLF.   In
particular, they found  that, for a given mass,  the luminosity of the
central galaxy of  a halo has a fairly  narrow distribution, while the
number  of satellite  galaxies roughly  obeys a  Poisson distribution,
consistent  with  the  subhalo  statistics  in  numerical  simulations
(Kravtsov \etal 2004).

Most of  the halo-occupation analyses mentioned above  focused on the
occupation numbers of  galaxies in dark matter haloes,  with little or
no attention to the details regarding how these galaxies are spatially
distributed  within  their  haloes.   In modeling  galaxy  correlation
functions on small scales, the  usual assumption is that the brightest
halo  galaxy resides  (at rest)  at the  halo center,  with  the other
galaxies   (hereafter   satellites)   following   a   number   density
distribution that is  identical to that of the  dark matter particles. 
Although this  assumption yields correlation functions  that match the
observations reasonably  well, the details  certainly have to  be more
complicated. For example, it is  well known that galaxies of different
types follow different spatial distributions in galaxy systems (e.g.
Dressler 1980; Postman \& Geller 1984;  Adami, Biviano \& Mazure 1998; 
Dominguez et al. 2001; Goto \etal 2003; Magliochetti \& Porciani 2003; 
Madgwick \etal 2003; Scranton 2003; Collister \& Lahav 2004). Red and 
early-type galaxies are preferentially found towards the centers of 
large groups. 

An  attractive method to  probe the  spatial distribution  of galaxies
with  respect to the  dark matter  haloes, is  to use  the galaxy-halo
cross-correlation  function.   Although  dark  matter haloes  are  not
directly observable,  one can  use galaxy groups  as a  surrogate (see
Yang \etal 2005a), and use the galaxy-group cross correlation function
instead. Since  this cross correlation  function is an average  of the
excess of galaxies  at a given distance from the  group center, it can
be interpreted as the average, radial distribution of galaxies both in
and around their dark matter haloes.   In what follows we will use the
terms  galaxy-halo cross correlation  function and  galaxy-group cross
correlation  function without  distinction, and  use  the abbreviation
GHCCF to indicate  either one.  With the large  and uniform catalogues
of galaxy  groups that can  be constructed from large  galaxy redshift
surveys, such as the  2dFGRS  and the  Sloan Digital Sky Survey (SDSS; 
York \etal 2000), the GHCCF can be studied as a function of group mass.  
Furthermore,   since  these   redshift  surveys   contain  information
regarding various detailed properties of the individual galaxies, such
as  luminosity, stellar  mass,  colour, spectral  type, star formation
rate, morphological  type, etc.,  the cross correlation  technique can
also be  used to  study the spatial  distributions of galaxies  in and
around dark matter haloes as function of these physical properties.

In this  paper, we  use large catalogues  of galaxy  groups, extracted
from the 2dFGRS  and SDSS using the halo-based  group finder developed
by Yang \etal (2005a), to study the GHCCF as a function of luminosity,
colour,  spectral  type,  and  specific  star  formation  rate  of  the
galaxies, and as a function of  group mass.  These results are used to
infer the spatial distributions of  different kinds of galaxies in and
around dark  matter haloes of  different masses.  The outline  of this
paper is  as follows.  In Section~\ref{sec_GC}, we  describe our group
and  galaxy  catalogues.    Section~\ref{sec_2pcf}  describes  how  we
determine the GHCCFs. Our results are presented in Section
~\ref{sec_results}, and we compare the 2dFGRS observation with 
realistic mock galaxy redshift surveys in Section~\ref{sec_mock}. 
Finally, we summarize our results in Section~\ref{sec_summary}.

\section{The Data}
\label{sec_GC}

\subsection{Group Selection}
\label{sec:groups}

In Yang \etal (2005a; hereafter YMBJ), we developed a halo-based group
finder that can successfully  assign galaxies into groups according to
their  common haloes.   The basic  idea  behind this  group finder  is
similar to that  of the matched filter algorithm  developed by Postman
\etal (1996),  although it  also makes use  of the galaxy  kinematics. 
The group finder starts with  an assumed mass-to-light ratio to assign
a  tentative  mass  to  each  potential group,  identified  using  the
friends-of-friends (FOF)  method.  This mass  is used to  estimate the
size and  velocity dispersion  of the underlying  halo that  hosts the
group,  which  in turn  is  used  to  determine group  membership  (in
redshift space).  This procedure  is iterated until no further changes
occur  in  group memberships.   Using  detailed  mock galaxy  redshift
surveys, the performance of our  group finder has been tested in terms
of completeness of true members and contamination by interlopers.  The
average  completeness of individual  groups is  $\sim 90$  percent and
with only  $\sim 20$ percent interlopers.   Furthermore, the resulting
group catalogue is insensitive to the initial assumption regarding the
mass-to-light ratios, and is more successful than the conventional FOF
method in  associating galaxies according to their  common dark matter
haloes.

In this paper  we use this group finder  to construct group catalogues
from both the 2dFGRS and the SDSS, which we briefly describe below.

\begin{table*}
\caption{Numbers of galaxies and groups in volume-limited samples.} 
\label{tab:number}
\begin{tabular}{lccc} 
\hline
 & $z<0.09$ & $z<0.13$ & $z<0.18$  \\
 & (1) & (2) & (3)  \\
\hline\hline
Groups & 2dF / SDSS & 2dF / SDSS & 2dF / SDSS \\
$12.2\le \log M_{\rm h}<13.0 $ & 4846 / 8994 & 14189 / 19615 & - / -  \\
$13.0\le \log M_{\rm h}<13.8 $ & 878 / 1445  & 2571 / 4237   & - / -  \\
$13.8\le \log M_{\rm h} $      & 129 / 202   & 382 / 574     & 977 / 1619 \\
\hline
Galaxies in 2dFGRS & early / late & early / late & early / late \\
$M_{b_J}-5\log h<-18.0$  & 13832 / 22754 & - / - & - / -   \\
$M_{b_J}-5\log h<-19.0$  & 7436 / 8013   & 23320 / 23792 &  - / -    \\
$M_{b_J}-5\log h<-20.0$  & 2043 / 1277   & 6914 / 3835 & 16403 / 10361      \\
\hline
Galaxies in SDSS & red / blue & red / blue & red / blue \\
$M_{r,0.1}-5\log h<-19.5$  & 23129 / 21626 & - / - & - / -      \\
$M_{r,0.1}-5\log h<-20.5$  & 7899 / 4650   & 21969 / 13811 & - / -     \\
$M_{r,0.1}-5\log h<-21.5$  & 837 / 185     & 2451 / 577    &  6583 / 1624   \\
\hline
\end{tabular}
\end{table*}

\subsection{The 2dFGRS}
\label{sec:2dfgrs}

We use the final, public  data release from the 2dFGRS, which contains
about  $250,000$  galaxies  with  redshifts  and  is  complete  to  an
extinction-corrected apparent magnitude of $b_J\approx 19.45$ (Colless
\etal 2001). The survey volume  of the 2dFGRS consists of two separate
declination  strips in  the North  Galactic Pole  (NGP) and  the South
Galactic Pole (SGP), respectively, together with 100 two-degree fields
spread  randomly  in  the   southern  Galactic  hemisphere.   For  the
construction  of our group  catalogue, we  restrict ourselves  only to
galaxies in  the NGP and SGP  regions, and with  redshifts $0.01\leq z
\leq  0.20$,  redshift  quality  parameter  $q \geq  3$  and  redshift
completeness $c>0.8$.  This leaves a grand total of $151,280$ galaxies
with  a  sky coverage  of  $1124 \,  {\rm  deg}^2$.   The typical  rms
redshift  and   magnitude  errors  are  $85  \kms$   and  $0.15$  mag,
respectively (Colless  \etal 2001).  Absolute  magnitudes for galaxies
in the 2dFGRS  are computed using the K-corrections  of Madgwick \etal
(2002).

Application  of the  halo-based group  finder to  this  galaxy sample,
yields  a group  catalogue consisting  of $77,708$  systems,  which in
total  contain $104,912$  galaxies.  Among  these systems,  $7251$ are
binaries,  $2343$  are triplets,  and  $2502$  are  systems with  four
members or more.   The vast majority of the  groups ($66,612$ systems)
in our catalogue, however, consist of only a single member.  Note that
some  faint galaxies  are not  assigned to  any group,  because  it is
difficult to decide whether they  are the satellite galaxies of larger
systems,  or   the  central   galaxies  of  small   haloes.   Detailed
information regarding the  clustering properties and galaxy occupation
statistics of these groups can be found in Yang \etal (2005a,b,c).

As discussed in YMBJ, it is not reliable to estimate the (total) group
luminosity based on the assumption that the galaxy luminosity function
in groups  is similar to  that of field  galaxies. We therefore  use a
more  empirical approach  to estimate  the group  luminosity $L_{18}$,
defined as  the total  luminosity of all  group members  brighter than
$M_{b_J}-5\log h  = -18$.  We refer  the reader to  Yang \etal (2005b)
for  details about  how  $L_{18}$  is estimated  for  each group.   

As demonstrated in detail in YMBJ, $L_{18}$ is tightly correlated with
the mass of the dark matter halo hosting the group, and can be used to
rank galaxy groups according to halo masses. To this extent we use the
mean group separation, $d=n^{-1/3}$, as a mass indicator.  Here $n$ is
the number density of all  groups brighter (in terms of $L_{18}$) than
the group in consideration.  Since $L_{18}$ is tightly correlated with
halo mass  $M$, we can  convert $d$ to  $M$ (see Yang \etal  2005c for
details).   Unfortunately, this conversion  requires knowledge  of the
halo mass function, and  is therefore cosmology dependent.  Throughout
this  paper we  consider a  $\Lambda$CDM `concordance'  cosmology with
$\Omega_m = 0.3$, $\Omega_{\Lambda}=0.7$, $h=0.7$ and $\sigma_8=0.9$.

\subsection{The SDSS}
\label{sec:sdss}

We have  also applied our group finder  to the SDSS.  Here  we use the
New   York   University   Value-Added  Galaxy   Catalogue   (NYU-VAGC)
\footnote{http://wassup.physics.nyu.edu/vagc/\#download},   which   is
described in detail in Blanton \etal (2005).  The NYU-VAGC is based on
the  SDSS  Data  Release  2   (Abazajian  \etal  2004),  but  with  an
independent  set  of  significantly  improved reductions.   From  this
catalogue we select all galaxies  in the Main Galaxy Sample, which has
an extinction corrected Petrosian magnitude limit of $r=18$.  We prune
this sample to those galaxies in  the redshift range $0.01 \leq z \leq
0.20$ and with a redshift completeness $c > 0.7$.  This leaves a grand
total of $184,425$ galaxies with a  sky coverage of $\sim 1950 \, {\rm
  deg}^2$.  For SDSS galaxies, we  also use stellar masses and current
star  formation  rates released  by  Brinchmann  et  al. (2004b).  The
stellar masses of individual  galaxies are estimated from the observed
stellar  absorption  indices (Kauffmann  et  al.  2003a,b), while  the
current star  formation rates of individual galaxies  are estimated by
fitting the observed spectra with spectral synthesis model (Brinchmann
et al. 2004a). For the SDSS  sample used in this paper, more than 90\%
of  the galaxies  have  estimated stellar  masses  and star  formation
rates.  We  use only  these galaxies to  form subsamples  according to
stellar mass or star formation rate. We have tested that the inclusion
of galaxies without stellar mass and star formation estimates does not
have a significant impact on our results.
 
From this  SDSS sample, we  construct a group catalogue  that contains
$102,935$ systems.   Among these systems, $9831$  are binaries, $3042$
are triplets,  $3473$ are systems with  four or more  members, and the
majority ($86,589$ systems) have only a single member. A more detailed
description  of this  catalogue will  be presented  in  Weinmann \etal
(2005,  in preparation).   As for  the 2dFGRS,  we estimate  the group
luminosity  $L_{195}$, defined as  the total  luminosity of  all group
members  brighter  than   $M_{r,0.1}-5\log  h  =  -19.5$\footnote{Here
  $M_{r,0.1}$ is the absolute magnitude in the $r$-band, $K$-corrected
  to  a redshift of  $0.1$ (see  Blanton \etal  2003a for  details).}. 
Finally, we  use the  rank of  $L_{195}$ to assign  each group  a halo
mass, using the same technique as described above for the 2dFGRS.

   It is interesting to note that there is an overlapping region 
for the 2dFGRS and SDSS near the North Galactic Pole. 
We found 457 SDSS groups containing $\ge 3$ galaxies, among which 
418 were also selected as 2dF groups. The SDSS and 2dF groups are not 
identical, because the selection effects are different for the two 
surveys, but their properties are similar.    

\begin{figure*}
\centerline{\psfig{figure=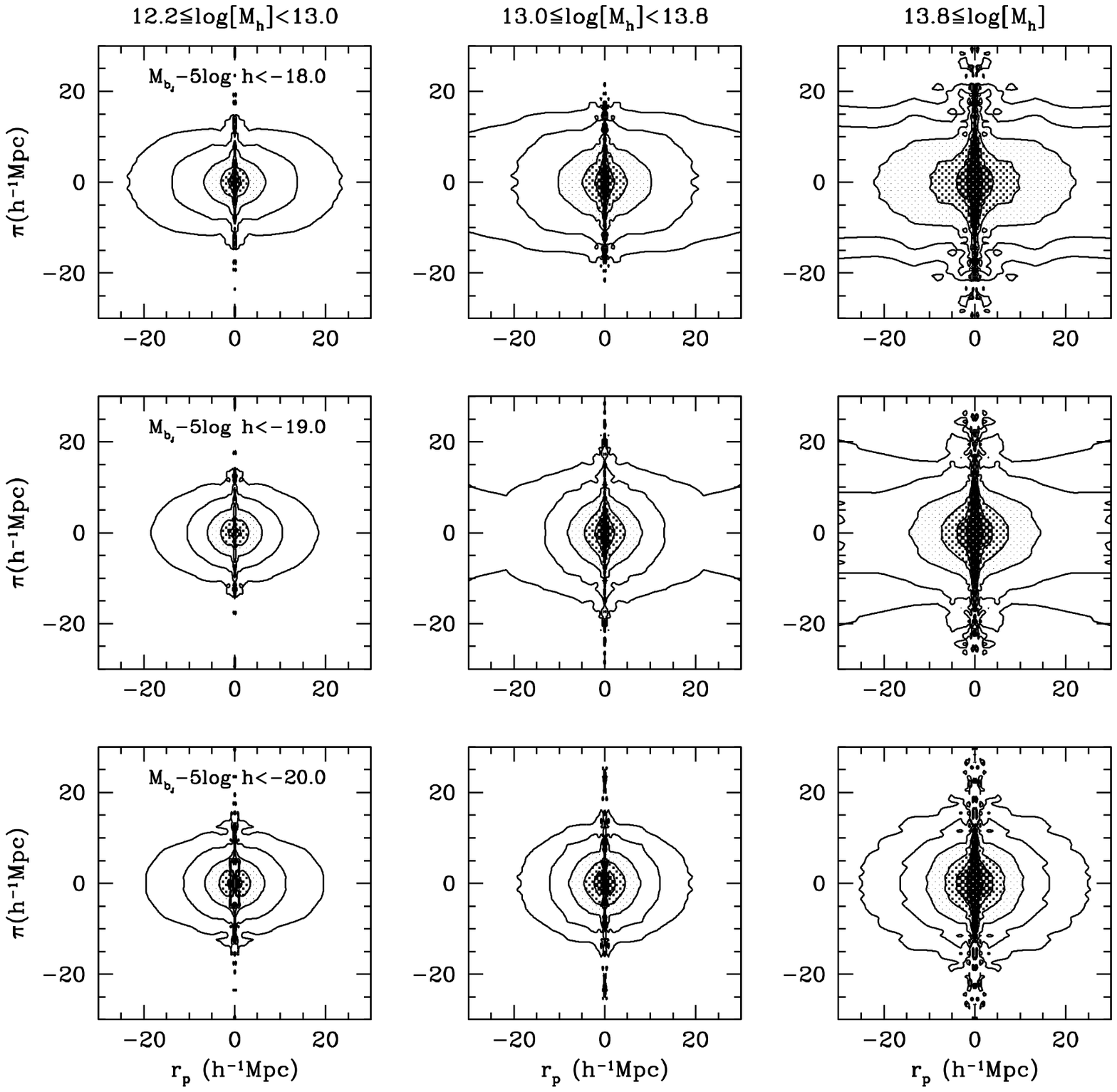,width=\hdsize}}
\caption{The cross-correlation function, $\xi(r_p,\pi)$, for
  various groups and galaxies  extracted from the 2dFGRS.  Panels from
  left to right  correspond to groups of different  halo masses, while
  panels  from top  to bottom  correspond to  galaxies  with different
  magnitude limits (as indicated).}
\label{fig:xi}
\end{figure*}

\section{The group-galaxy cross correlation function}
\label{sec_2pcf}

In redshift space, the separation  between a group center and a galaxy
can be  split in the  separations perpendicular, $r_p$,  and parallel,
$\pi$, to  the line-of-sight. Explicitly, for a  pair $({\bf s_1},{\bf
  s_2})$, with ${\bf s_i} = c z_i {\bf \hat{r}_i}/H_0$, we define
\begin{equation}
\label{rppi}
\pi = {{\bf s} \cdot {\bf l} \over \vert {\bf l} \vert} \; ,
\;\;\;\;\;\;\;\;\;\;\;\;\;\;
r_p = \sqrt{{\bf s} \cdot {\bf s} - \pi^2}
\end{equation}
Here ${\bf l}={1\over2}({\bf  s_1} + {\bf s_2})$ is  the line of sight
intersecting the  pair, and  ${\bf s}  = {\bf s_1}  - {\bf  s_2}$.  We
compute the galaxy-group  (or galaxy-halo) two-point cross correlation
function (GHCCF), $\xi(r_p,\pi)$, using the following estimator
\begin{equation}
\label{tpcfest}
\xi(r_p,\pi) = {N_R \over N_D} 
{\langle GD \rangle \over \langle GR \rangle} - 1
\end{equation}
where  $N_D$, $N_R$  are the  number  of galaxies  and random  points,
respectively, and  $\langle GD \rangle$  and $\langle GR  \rangle$ are
the  number of  group-galaxy  and group-random  pairs with  separation
$(r_p,\pi)$. Each galaxy  (random  point) is  weighted  by the inverse  
of the survey redshift completeness $c_i$.

Throughout this paper, we use volume-limited samples for both galaxies
and groups. As discussed in Yang \etal (2005c), groups with given halo
masses  are   complete  only  to   a  certain  redshift. 
To ensure completeness,  we  use systems at $z \le  0.13$ 
for  haloes with masses in the range 
$12.2 \le  \log M_{h}/(h^{-1}M_{\odot})<13.8 $,
and systems at $z \le 0.18$ for haloes with masses   
in the range $13.8 \le \log M_{h}/(h^{-1}M_{\odot}) $.
For the galaxies,  we consider three  volume-limited samples  corresponding to
the following three redshift limits: $z=0.09$, $0.13$, $0.18$. For the
2dFGRS, these redshift  limits correspond to absolute-magnitude limits
$M_{b_J}-5 \log  h= -18.0$, $-19.0$, and  $-20.0$, respectively, while
for the SDSS, they correspond to $M_{r,0.1}-5 \log h= -19.5$, $-20.5$,
and  $-21.5$.  In Table~\ref{tab:number}, we list the
number of groups and the number of galaxies in each of these 
volume-limited samples.
When  measuring the  GHCCFs, we  restrict  galaxies and
groups to the redshift range in which both the groups and galaxies are
complete. To  normalize the  correlation function defined  in equation
(\ref{tpcfest}), we generate a random sample that is 50 times as large
as the corresponding real sample (i.e. $N_R=50 N_D$).

The separations $r_p$ and $\pi$  are defined with respect to the group
centers.  Since galaxy groups have non-negligible sizes, the GHCCF can
depend  sensitively on  how  exactly the  group  centers are  defined,
especially on small scales. To probe this sensitivity, and, as we will
show,   to  gain   valuable  insights,   we  consider   two  different
definitions: the luminosity-weighted coordinates of the group members,
and the location of the brightest galaxy in the group. In what follows
we refer to  these as the LW and BG  centers, respectively.  Note that
these two definitions may give quite different results, especially for
small groups with only a  few members.  For instance, consider a group
with  only  two  members  of  comparable luminosity,  separated  by  a
distance $r$.   The LW center will  be roughly midway  between the two
galaxies, while the  BG center is located at one of  the two galaxies. 
This leads  to strong differences  in the two-point  cross correlation
function.  In the  first case, there are two  group-galaxy pairs, both
with separations  $\sim r/2$.  In the  second case, there  is only one
group-galaxy pair  with a separation  $r$; by definition,  the central
galaxy is  at zero distance from  the group center  and so contributes
only to the correlation function at the zero lag.

The  LW  and  BG  centers  have  different  physical  motivations  and
interpretations.  If  light traces mass,  at least within  dark matter
haloes, the LW centers seem  a natural choice. However, because of the
discreteness of the galaxies, it is clear that, even when light traces
mass  accurately in  a statistical  sense,  it is  not necessarily  an
accurate description in  individual systems with only a  few galaxies. 
The  BG  centers are  motivated  by  the  standard picture  of  galaxy
formation,  according to  which  the  brightest galaxy  in  a halo  is
expected to reside at rest at  the halo center.  If this is indeed the
case, the BG  centers are clearly a very physical  and natural choice. 
Furthermore, this definition does  not suffer from the discreteness of
galaxies, but instead is  based on it.  Unfortunately, as demonstrated
in  van den Bosch  \etal (2005b),  there is  strong evidence  that, on
average, the brightest  halo galaxy has a significant  offset from the
center of  the dark matter halo.   This most likely  reflects that the
majority of dark matter haloes is not yet fully relaxed, implying that
there is  no well defined center  at all. Note that  this ambiguity in
defining halo centers  exists also for dark matter  haloes in $N$-body
simulations, where the center of mass and the minimum of the potential
often do not coincide. Given these difficulties, we feel that the best
approach is  simply to  use both definitions,  and to see  whether the
differences in the resulting GHCCFs can provide new insights.

Fig.~\ref{fig:xi}  shows  the contours  of  $\xi(r_p,\pi)$ for  groups
(haloes)   of  different   masses  and   for  galaxies   of  different
luminosities.  For conciseness, we only  show the results based on the
LW centers.  Note that here and in the following we use volume-limited
samples both  for galaxies and  galaxy groups. The effect  of redshift
space distortions  is clearly visible: on  small scales $\xi(r_p,\pi)$
is stretched  in the $\pi$-direction, due to  the peculiar, virialized
motions of galaxies within dark  matter haloes. Note that this effect,
often called the Finger-of-God effect,  is much more pronounced in the
more  massive  haloes (right-hand  columns),  reflecting their  larger
velocity dispersions. On large scales, the contours are squashed along
the  line-of-sight direction, due  to the  infall effect  discussed by
Kaiser (1987).  In  a separate paper (Li \etal  2005, in preparation),
we model these redshift space  distortions in detail, in order to infer
the velocity  field and mass  density distribution in and  around dark
haloes.   In  what  follows,  we  only  focus  on  the  projection  of
$\xi(r_p,\pi)$ along the $\pi$-direction (i.e., the line-of-sight):
\begin{equation}
\label{project}
w_p(r_p) = \int_{-\infty}^{\infty} \xi(r_p,\pi) {\rm d}\pi = 
2 \int_{r_p}^{\infty} \xi(r) \, {r \, {\rm d}r \over \sqrt{r^2 - r_p^2}}
\end{equation}
The second equality  shows that $w_p(r_p)$ is a  simple Abel transform
of the real-space cross correlation function, $\xi(r)$. This owes to the
 overall  isotropy and to the fact that the redshift-space distortions
only affect  $\pi$, but  not $r_p$, and  implies that  $w_p(r_p)$ will
have a power-law shape as long  as $\xi(r)$ has a power-law shape.  In
practice,  we  integrate   equation  (\ref{project})  over  the  range
$|\pi|\le 40 \mpch$.  From now on,  whenever we refer to the GHCCF, we
mean this {\it projected} cross-correlation function $w_p(r_p)$.

\begin{figure*}
\centerline{\psfig{figure=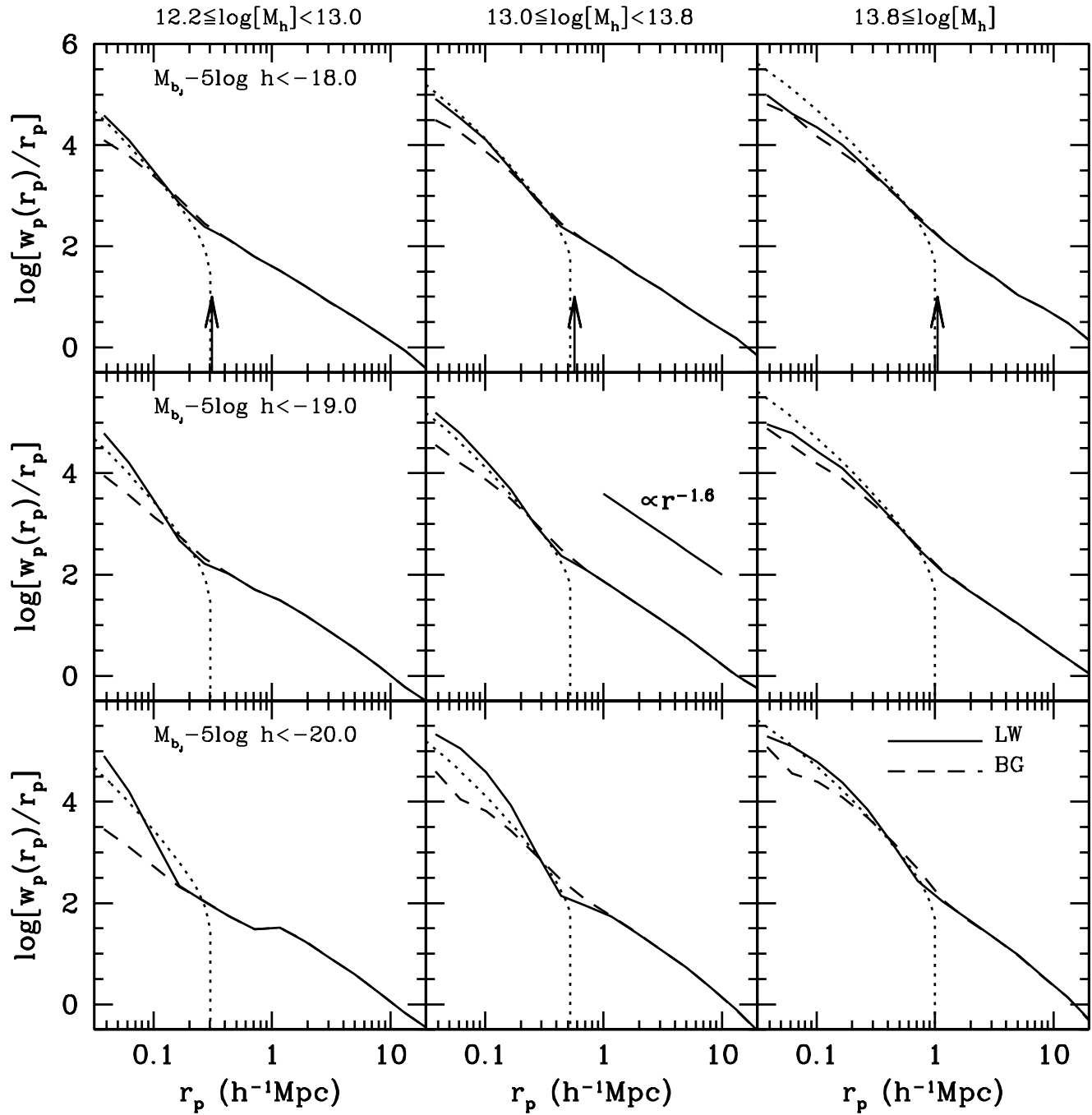,width=\hdsize}}
\caption{The projected cross-correlation function between galaxies and
  groups  in  the  2dFGRS.   Solid  and  dashed  lines  correspond  to
  luminosity weighted  (LW) and  brightest galaxy (BG)  group centers,
  respectively.    The  dotted  lines   illustrate  the   1-halo  term
  corresponding to an  NFW profile that belongs to a  halo with a mass
  that is equal to the median mass of the range considered. The arrows
  in the  upper panels indicate  the corresponding virial  radii.}
\label{fig:wrp}
\end{figure*}
\begin{figure*}
\centerline{\psfig{figure=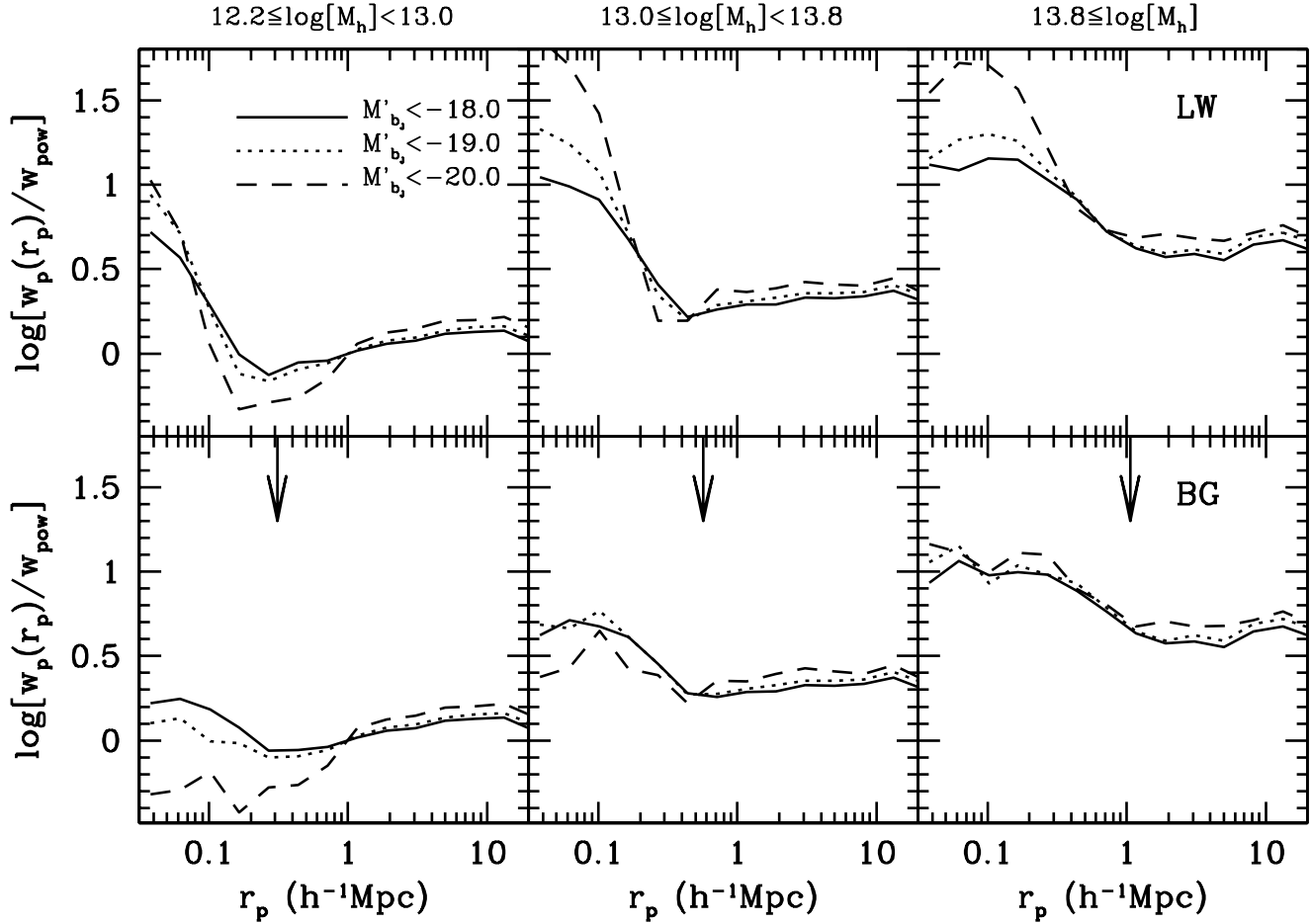,width=\hdsize}}
\caption{The ratio of the projected 2dFGRS cross-correlation function and 
  the power-law  relation of equation~(\ref{power_law}).   Results are
  shown for three lower limits  on the galaxy luminosity, as indicated
  (with $M'_{b_J}=M_{b_J} -  5 {\rm log} h$).  Upper  and lower panels
  show  the  results  obtained  using  luminosity  weighted  (LW)  and
  brightest galaxy  (BG) group centers,  respectively. Arrows indicate
  the virial radii of the median-mass haloes (cf. Fig.~\ref{fig:wrp}).
  Note that  in order to eliminate  the impact of  cosmic variance, we
  restrict all galaxy and group  samples to the same volume with $0.01
  \le z \le 0.09$ (see text for discussion).}
\label{fig:ratio_lum}
\end{figure*}
\begin{figure*}
\centerline{\psfig{figure=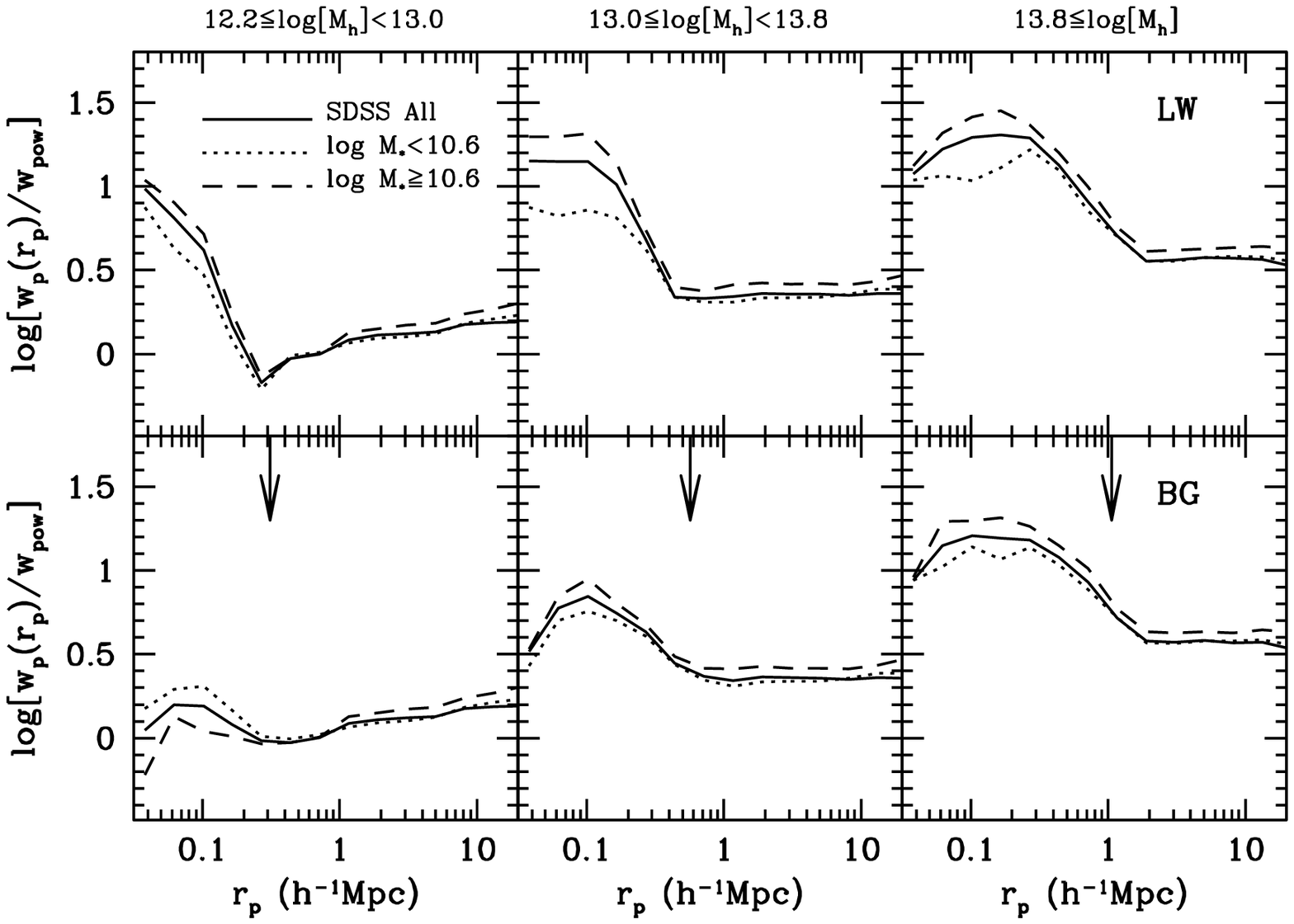,width=\hdsize}}
\caption{Similar to Fig~\ref{fig:ratio_lum}, but for SDSS galaxies 
  of  different  stellar masses.  Note  that  these  samples are  also
  restricted to redshifts $0.01 \le z \le 0.09$.}
\label{fig:ratio_M*}
\end{figure*}

\section{Results}
\label{sec_results}

\subsection{The shape of the cross correlation function}
\label{sec:shape}

Fig.\,\ref{fig:wrp}  shows  the  GHCCF  between  galaxies  and  groups
obtained  from the  2dFGRS.   The  solid and  dashed  curves show  the
results obtained using the LW and BG centers, respectively. When using
the  LW centers,  the GHCCFs  clearly reveal  two distinct  regimes, a
steep inner  part and  a relatively flat  outer part.   The transition
between these two regimes occurs at a radius that is comparable to the
virial radius of  the haloes in consideration (indicated  by arrows in
the  upper  panels, and  described  in  more  detail below).   In  the
terminology of the halo model  (e.g., Cooray \& Sheth 2002), the inner
part of the GHCCF is dominated by the `1-halo' term, in the sense that
the galaxy-group pairs are dominated  by the ones between galaxies and
their own host group, while on larger scale, the GHCCF is dominated by
pairs between groups  and the member galaxies of  other groups  or  of  
galaxies not in groups.  When
using  the  BG  centers,   the  small  scale  GHCCF  is  significantly
shallower,  with no  clear transition  from the  `1-halo'  to `2-halo'
regimes. The difference between the small scale GHCCFs based on the LW
and BG  centers is  most pronounced in  the cross  correlation between
low-mass groups and bright  galaxies.  This suggests that these groups
typically contain only  a single bright galaxy near  their LW centers. 
In this case, the number  of group center/galaxy pairs on small scales
is greatly  reduced if the central  galaxies are not used  in the pair
count, as in the case of the BG center.
 
For  pair  separations  larger  than  the  virial  radius,  the  exact
definition of  the group center is  not important, and  all GHCCFs are
roughly parallel to  each other, independent of the  group mass or the
galaxy luminosity  function.  To good approximation,  this large scale
GHCCFs can be  described by a power law  $w(r_p)/r_p \propto r^{-1.6}$
(indicated   by   a   straight   line   in  the   central   panel   of
Fig.\,\ref{fig:wrp}).  This is in  agreement with the linear halo bias
model (Mo \& White 1996),  which states that at large (linear) scales,
the real-space  cross correlation function between haloes  of mass $M$
and galaxies of luminosity $L$ can be written as
\begin{equation}
\label{linbias}
\xi_{gh}(r) = b_g(L) b_h(M) \xi_{\rm dm}(r)\,, 
\end{equation}
with  $\xi_{\rm  dm}(r)$ the  dark  matter  correlation function,  and
$b_g(L)$ and  $b_h(M)$ the bias of  galaxies of luminosity  $L$ and of
haloes of  mass $M$, respectively. As  long as this  linear bias model
applies,  it   is  therefore  expected  that   the  galaxy-halo  cross
correlation  functions all have  the same  form, with  a normalization
that depends on the luminosities and masses of the galaxies and haloes
considered.

The   dotted  curves   in  Fig.\,\ref{fig:wrp}   show   the  projected
correlation function  obtained for NFW (Navarro, Frenk  \& White 1997)
profiles of dark matter particles. For each of the three mass bins, we
use  the  mean mass  of  the groups  in  consideration  to estimate  a
`virial' radius [marked as arrows;  defined as the radius within which
the mean overdensity  is that given by the  spherical collapse model],
and to obtain  a halo concentration parameter using  the model of Eke,
Navarro \& Steinmetz (2001). The profile is assumed to be truncated at
the  `virial'  radius,  and  the  projected  correlation  function  is
obtained by integrating the NFW profile along the line-of-sight. Using
the LW group centers, and including faint galaxies in the group-galaxy
pair  counts, yields  a GHCCF  that roughly  follows the  NFW profile,
except for the most massive  groups where the NFW profile overpredicts
the actual GHCCF. In the case  of the BG centers, the `1-halo' part of
the GHCCF is much shallower than the NFW profile, especially for small
groups.   Although  it is  tempting  to  use  this NFW  comparison  to
constrain the spatial  bias of galaxies within dark  matter haloes, we
will demonstrate in Section~\ref{sec_mock} that this comparison is not
straightforward.  In particular,  using realistic Mock Galaxy Redshift
Surveys (MGRSs), we  will show that the GHCCF based  on the LW centers
does not reveal the {\it actual} distribution of galaxies.  Therefore, 
the  NFW comparison shown  here has to be  interpreted with care.

We have performed a similar analysis for the SDSS groups.  Since these
results are very similar to those  based on the 2dFGRS groups shown in
Figs.~\ref{fig:xi} and~\ref{fig:wrp}, we do not show them here.

\begin{figure*}
\centerline{\psfig{figure=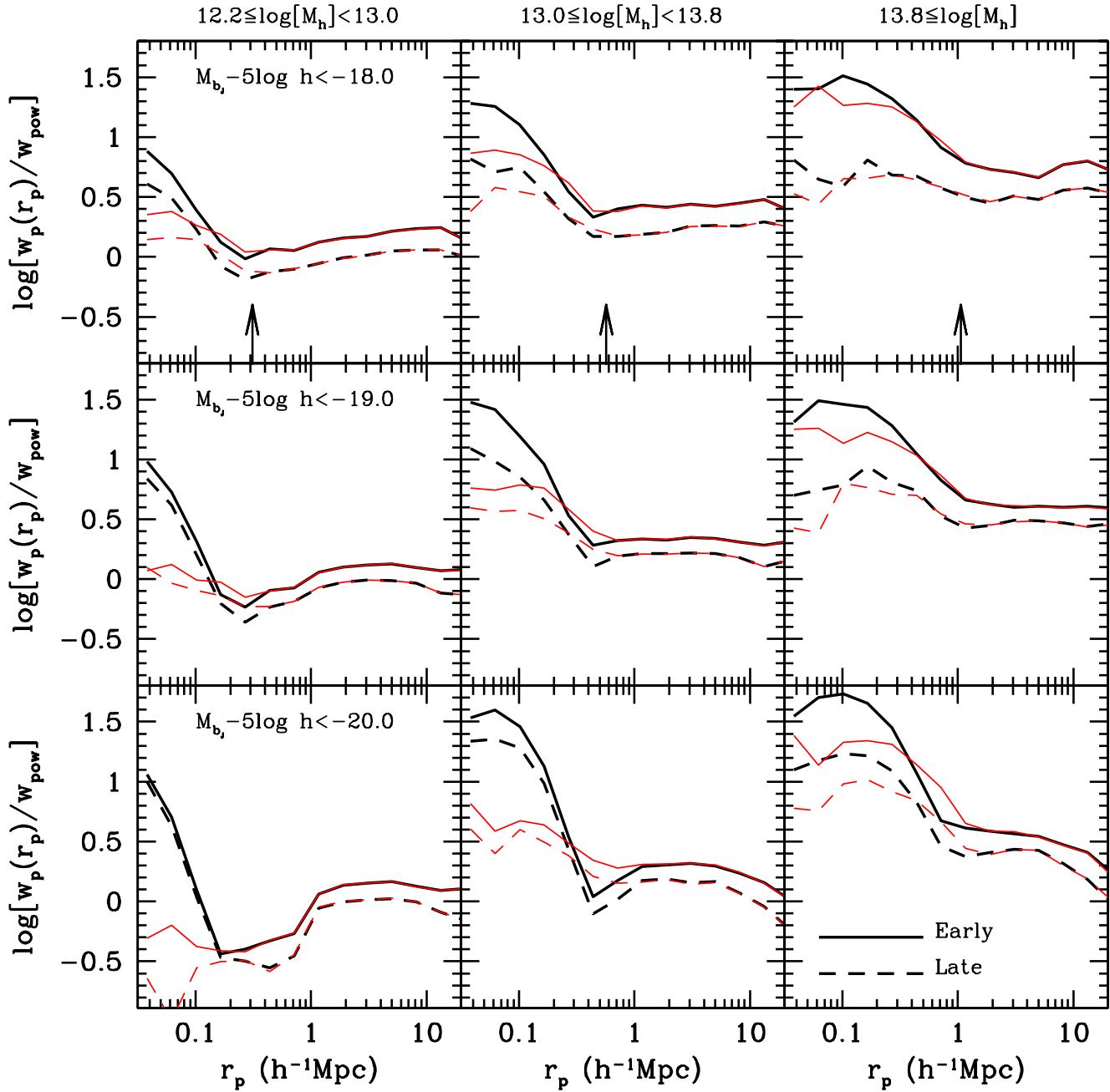,width=\hdsize}}
\caption{The ratio of the projected 2dFGRS cross-correlation function and 
  the power-law relation  of equation~(\ref{power_law}) for early-type
  (solid lines) and late-type (dashed lines) galaxies.  Thick and thin
  lines correspond to LW and BG centers, respectively.}
\label{fig:ratio_type}
\end{figure*}
\begin{figure*}
\centerline{\psfig{figure=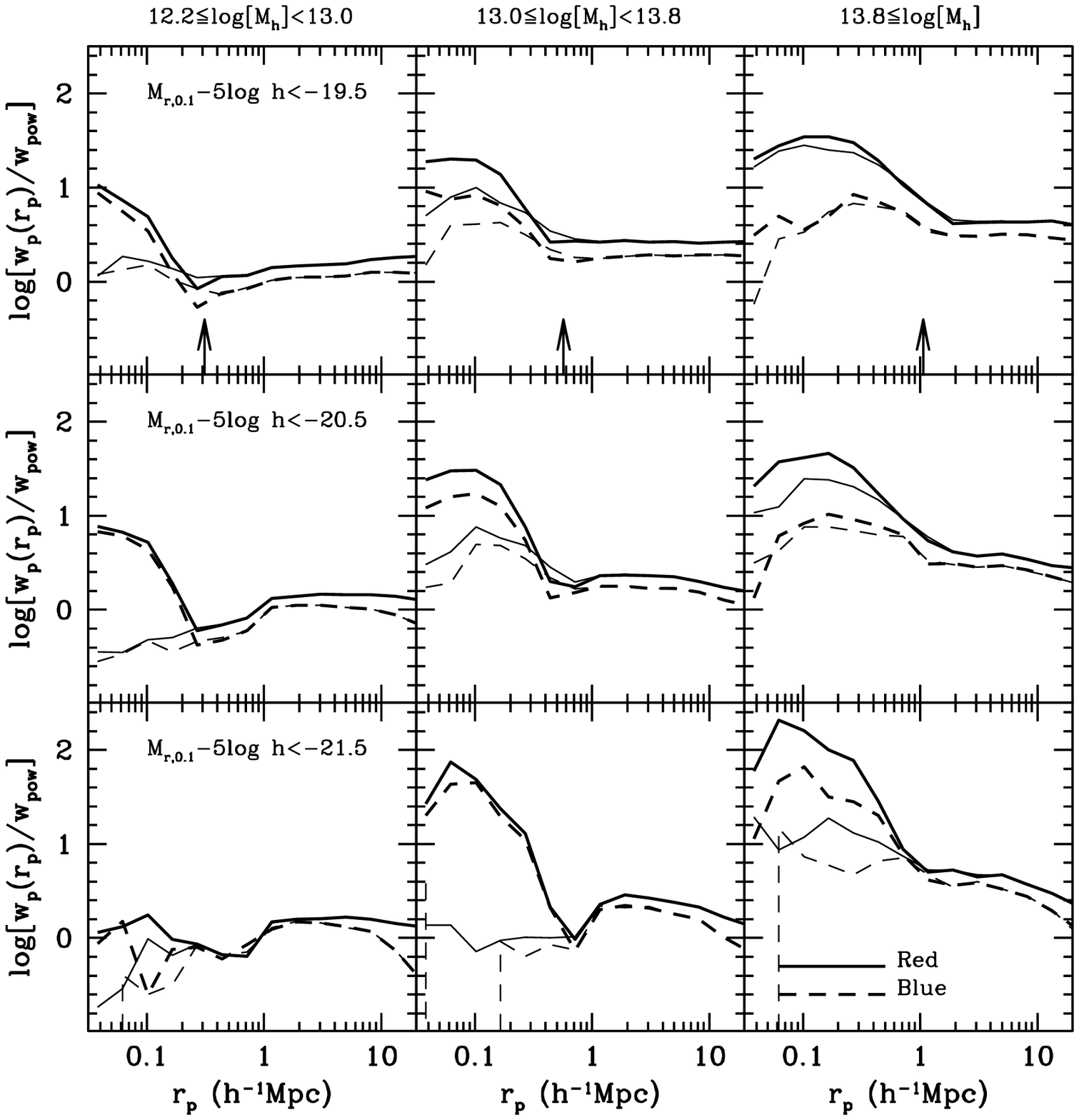,width=\hdsize}}
\caption{The ratio of the projected cross-correlation function and 
  the power-law relation  of equation~(\ref{power_law}) for red (solid
  lines) and blue (dashed lines) galaxies in the SDSS.  Thick and thin
  lines correspond to LW and BG centers, respectively.}
\label{fig:ratio_colour}
\end{figure*}

\subsection{Dependence on galaxy luminosity and stellar mass} 
\label{sec:stelmass}

Before  we proceed  to probe  the dependence  of the  GHCCF  on galaxy
luminosity  and stellar  mass, we  apply further  restrictions  to our
samples. The  2dFGRS contains two gigantic super-clusters;  one in the
NGP at $z\sim  0.08$, and the other in the SGP  at $z\sim 0.11$ (Baugh
\etal 2004). The presence of such structures can affect the clustering
statistics,  and so  care  must  be taken  when  comparing samples  of
different  depths.   In  order  to  eliminate  the   impact  of  these
extraordinary structures  on our investigation, in  this subsection we
restrict all  the galaxy and group  samples to the  same volume (i.e.,
all cut to redshift $z\le  0.09$), and make comparisons for samples in
the same  volume. The choice  of the cut  at $z=0.09$ is  a compromise
between having  large volume and having completeness  for the faintest
galaxies in consideration ($M_{b_j}-5\log(h)=-18$). Note that in order
to have sufficient large volume for good statistics, this redshift cut 
still includes the supercluster at $z=0.08$.

Fig.\,\ref{fig:ratio_lum}  shows  how  the  GHCCF  depends  on  galaxy
luminosity. Here we plot the ratio between the GHCCF and the following
power law:
\begin{equation}
\label{power_law}
w_p(r_p)/r_p = \left({r_p \over 10 h^{-1}\Mpc}\right)^{-1.6}
\end{equation}
As shown in Fig.\,\ref{fig:wrp}, this  power law matches the shapes of
all GHCCFs on large scales.  The amplitudes of the large scale GHCCFs,
however, are different for different group masses and different galaxy
luminosities,  reflecting the  mass and  luminosity dependence  of the
halo and  galaxy bias, respectively (cf.,  eq.[\ref{linbias}]).  As is
easily inferred from Fig.\,\ref{fig:wrp}, the halo bias $b_h(M)$ is an
increasing function of halo mass, while the galaxy bias $b_g(L)$ is an
increasing  function  of luminosity.   This  immediately implies  that
brighter  galaxies are  preferentially found  in more  massive haloes,
which is  the principle on which the  conditional luminosity formalism
is based (Yang \etal 2003, 2005c).

The  behavior on  scales smaller  than  the virial  radius (i.e.   the
`1-halo'  term)  is more  complicated,  and  depends  strongly on  the
definition of the group center.  In  the case of the LW centers (upper
panels),  bright galaxies have  a steeper  cross correlation  on small
scales  than fainter  galaxies. This  suggests that  brighter galaxies
have  a  more  concentrated  radial distribution  within  dark  matter
haloes.  However, if one uses  the BG centers (bottom panels), so that
the brightest galaxies themselves are not included in the group-galaxy
pair  counts, there  is no  significant luminosity  dependence  of the
GHCCF  in haloes with  $M\ga 10^{13}h^{-1}{\rm  M}_\odot$. This  is an
interesting  result, because  it  implies that  the strong  luminosity
segregation of galaxies observed in rich groups and clusters is almost
entirely  due  to the  brightest,  central  galaxy  in the  halo;  any
luminosity segregation  of satellite galaxies  in these systems  is at
best weak.  For haloes  with masses $M \la 10^{13}h^{-1}{\rm M}_\odot$
the small scale GHCCF is  in fact stronger for fainter galaxies.  This
suggests  that satellite  galaxies in  a low-mass  halo  are typically
significant  fainter than  their  brightest, central  galaxy, so  that
there are only a few pairs  of bright galaxies in these haloes.  As we
will show in Section~\ref{sec_mock}, similar results are obtained from
the MGRSs.

All  results  presented  above   are  based  on  luminosities  in  the
photometric $b_{J}$-band.   Since the mass-to-light ratio  of a galaxy
in this  (blue) band depends  strongly on its star  formation history,
the  lack of  luminosity segregation  of satellite  galaxies  does not
necessarily mean  a lack in  mass segregation. To test this, we now
turn to our SDSS group catalogue, where for each galaxy we also have 
estimates of their stellar mass and specific star formation rate
(see Section~\ref{sec:sdss}).

Fig.~\ref{fig:ratio_M*}  shows  the  GHCCFs  obtained  from  the  SDSS
groups, split into two subsamples according to the stellar mass of the
galaxies. The  dividing mass of $\log  M_{\star} = 10.6$  is chosen so
that the two  subsamples contain roughly the same  number of galaxies. 
Comparing the mass dependence  with the luminosity dependence shown in
Fig.~\ref{fig:ratio_lum}   one   notices   an   overall   resemblance,
indicating that, to  first order, more luminous galaxies  (in the blue
$b_J$-band)  are more  massive. However,  there are  also  some subtle
differences.  For  example, whereas the  small scale GHCCFs  reveal no
luminosity dependence for high mass  haloes when using the BG centers,
a  small stellar  mass  dependence is  apparent.   This suggests  that
satellite galaxies are  mildly segregated by mass. The  fact that this
effect is not seen when  using the $b_J$-band luminosities may reflect
that lower  mass galaxies  are relatively bluer.  Indeed, as  shown in
Kauffmann \etal  (2004), the specific star formation  rate and stellar
mass are anti-correlated.

Note also  that the  GHCCFs for  low mass haloes  on small  scales are
lower for more massive (in  stellar mass) satellite galaxies, which is
similar     to     the      luminosity     dependence     shown     in
Fig.\,\ref{fig:ratio_lum}.

\begin{figure}
\centerline{\psfig{figure=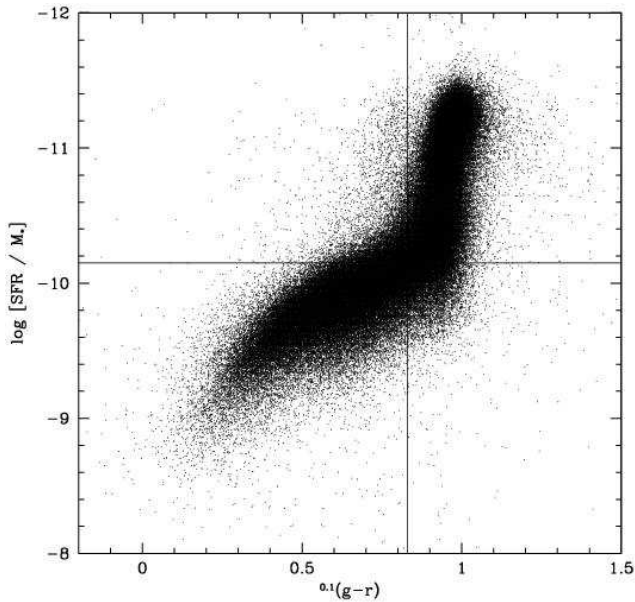,width=\hssize}}
\caption{ The colour-SSFR (specific star formation rate) relation 
  for SDSS galaxies used in our analysis.
  Vertical line is the dividing line we adopted to 
  separate galaxies into blue and red populations, while 
  the horizontal line is the dividing line we adopted 
  to separate galaxies into high- and low-SSFR populations. 
  Note that these two criteria separate galaxies in 
  approximately the same way.}
\label{fig:colour_SSFR}
\end{figure}

\subsection{Dependence on galaxy type, colour and star formation activity}
\label{sec:type}

Madgwick \etal  (2002) used a  principal component analysis  of galaxy
spectra   taken  from   the  2dFGRS   to  obtain   a   {\it  spectral}
classification  scheme  for  the   2dFGRS  galaxies.   They  used  the
parameter  $\eta$, a linear  combination of  the two  most significant
principal  components, to  classify galaxies  into  different spectral
types.  As  shown by Madgwick  \etal (2002), $\eta$ follows  a bimodal
distribution and can be interpreted  as a measure for the current star
formation rate in each  galaxy. Furthermore, $\eta$ is well correlated
with  {\it morphological} type  (Madgwick 2002).   In what  follows we
adopt  the classification  suggested  by Madgwick  \etal and  classify
galaxies with $\eta  < -1.4$ as `early-types' and  galaxies with $\eta
\geq -1.4$ as `late-types'.

Note that in all what follows, since we are not comparing the GHCCFs 
in different volumes, we do not restrict the galaxy and group  samples 
to the  same volume (i.e., the cut to redshift $z\le  0.09$).   
In  Fig.\,\ref{fig:ratio_type}, we  plot the  GHCCF obtained  from the
2dFGRS, divided by the the  power law (\ref{power_law}), for early and
late  type galaxies.   In all  cases, the  early-type galaxies  have a
larger correlation amplitude at scales  larger than the virial radius. 
Since  early-type galaxies  are preferentially  found in  more massive
haloes  and are, on  average, brighter  than late-type  galaxies, this
simply  reflects the fact  that $b_h(M)$  and $b_g(L)$  are increasing
functions of mass and luminosity, respectively.  In haloes with masses
$M \ga 10^{13}h^{-1}{\rm M}_\odot$ early-type galaxies tend to be more
centrally concentrated, as  is evident from the fact  that their GHCCF
on scales smaller than the halo virial radius is stronger.

This  is consistent  with Collister  \& Lahav  (2004), who  found that
early-type  galaxies in the  2dFGRS have  a more  concentrated profile
than late-type  galaxies, and dominate  the number counts  towards the
group (halo)  center.  However, we find that in lower mass haloes such  
a trend is much weaker.

By comparing  the GHCCFs  obtained from the  two definitions  of group
centers,  one can  see  that the  type  segregation in  haloes in  the
intermediate  mass range  is mainly  caused by  the central  galaxies. 
This  suggests  that in  haloes  with  masses $\sim  10^{13}h^{-1}{\rm
  M}_\odot$, early-type  galaxies start to dominate  the population of
central galaxies.  In the more massive haloes, early types continue to
have a  more concentrated distribution,  even if central  galaxies are
not  taken into account  (i.e., when  the BG  centers are  used). This
reflects the fact that massive haloes are dominated by early-types.

As mentioned  above, the spectral parameter $\eta$  can be interpreted
as a measure  for the current star formation activity  in each galaxy. 
Given  that star formation  activity is  strongly correlated  with the
optical colour of a galaxy, one expects to obtain similar results when
splitting  the sample  of galaxies  according to  colour,  rather than
according  to the  value of  $\eta$. To  test this,  we use  the $g-r$
colours of galaxies in the SDSS.  We split our sample of SDSS galaxies
into two  subsamples of roughly  {\it equal size}, by  using ${^{0.1}}
(M_g-M_r)=0.83$ as a dividing line. For the volume-limited sample used
in  our analysis,  this dividing  line is  approximately the  one that
separates  the bimodal  ${^{0.1}} (M_g-M_r)$  colour  distribution. In
what follows we refer to galaxies with ${ ^{0.1}}(M_g-M_r) > 0.83$ and
${  ^{0.1}}(M_g-M_r)<0.83$ as  red  and blue  galaxies, respectively.  
Note  that  since  we  are  only  interested  in  the  relative  color
dependence of  the GHCCFs,  we do  take account of  the fact  that the
bimodality  of the  galaxy color  distribution is  magnitude dependent
(e.g.,  Blanton \etal  2003b; Baldry  \etal  2004; Hogg  \etal 2004).  
Fig.\,\ref{fig:ratio_colour} plots the  GHCCFs between the SDSS groups
and these  two subsamples of  galaxies.  Comparing these  results with
those shown  in Fig.\,\ref{fig:ratio_type}, we see  that, as expected,
the  colour   dependence  of  the   GHCCF  is  very  similar   to  the
spectral-type dependence obtained from the 2dFGRS. 
There are noticeable differences for the brightest samples.
These differences are largely caused by the fact that
the luminosity cut is higher in the brightest sample 
shown in Fig.\,\ref{fig:ratio_colour} than that shown in 
Fig.\,\ref{fig:ratio_type}.

 We have also examined the dependence of the GHCCF on the {\it specific} star
formation  rate (SSFR),  which  we  defined as  the  ratio between  the
current star formation rate and  the stellar mass.  For the SDSS data,
both  these numbers are  obtained from  Kauffmann \etal  (2003a,b) and
Brinchman \etal (2004), as discussed in Section~\ref{sec:sdss}. We used
${\log \rm SSFR}=-10.15$ to  separate galaxies into high- and low-SSFR
subsamples  with similar  galaxy  numbers.  The  resulting GHCCFs  for 
these  subsamples are almost identical to those  shown in 
Fig.~\ref{fig:ratio_colour} based on colour separation. 
To better understand the similarity in the dependence of 
GHCCF on SSFR and colour,  we plot the SSFR-colour 
relation in Fig.\,\ref{fig:colour_SSFR}. Clearly, the separation 
at ${\log \rm SSFR}=-10.15$ and the separation at  
${  ^{0.1}}(M_g-M_r)=0.83$ lead to very similar subsamples. 
If  we separate galaxies according to their
{\it  absolute} star  formation  rate (SFR),  rather  than their  {\it
  specific}  star formation  rate (SSFR),  the difference  between the
high and low  SFR galaxies is much weaker. This owes  to the fact that
massive galaxies  in clusters may  still have considerable  amounts of
ongoing star formation, even though their SSFR is low.

The results obtained above show  that galaxies with 
low SSFRs (or red colours) are more
concentrated in  massive haloes than  galaxies with high  SSFRs
(bluer colours).  This
may  be interpreted  as evidence  that the  SFR is  suppressed  once a
galaxy comes  close to the center  of a massive  halo (cluster), where
the  interstellar gas  is striped  by  the hot  intra-cluster medium.  
However, the explanation is not  unique. It is also possible that more
massive galaxies have lower SSFRs (due to , e.g., stronger 
AGN feedback), and that they are more concentrated
because of dynamical friction. 
In this case, it is not the environment
but the stellar mass that determines the SSFR of a galaxy.  As we show
in  detail in  Weinmann \etal  (2005, in  preparation),  galaxies with
larger stellar masses have, on average, lower SSFRs. This implies that
the more concentrated distribution of  galaxies with a low SSFR should
at least  partially be due to  mass segregation. However,  for a given
stellar mass, there is also  a dependence on environment, in the sense
that the  fraction of  galaxies with low  SSFRs increases as  one goes
from low-mass  to high-mass  systems, or from  the outer to  the inner
regions in massive haloes (Weinmann \etal 2005, in preparation).  This
suggests  that environmental effects  may also  play some  role.  

\begin{figure*}
\centerline{\psfig{figure=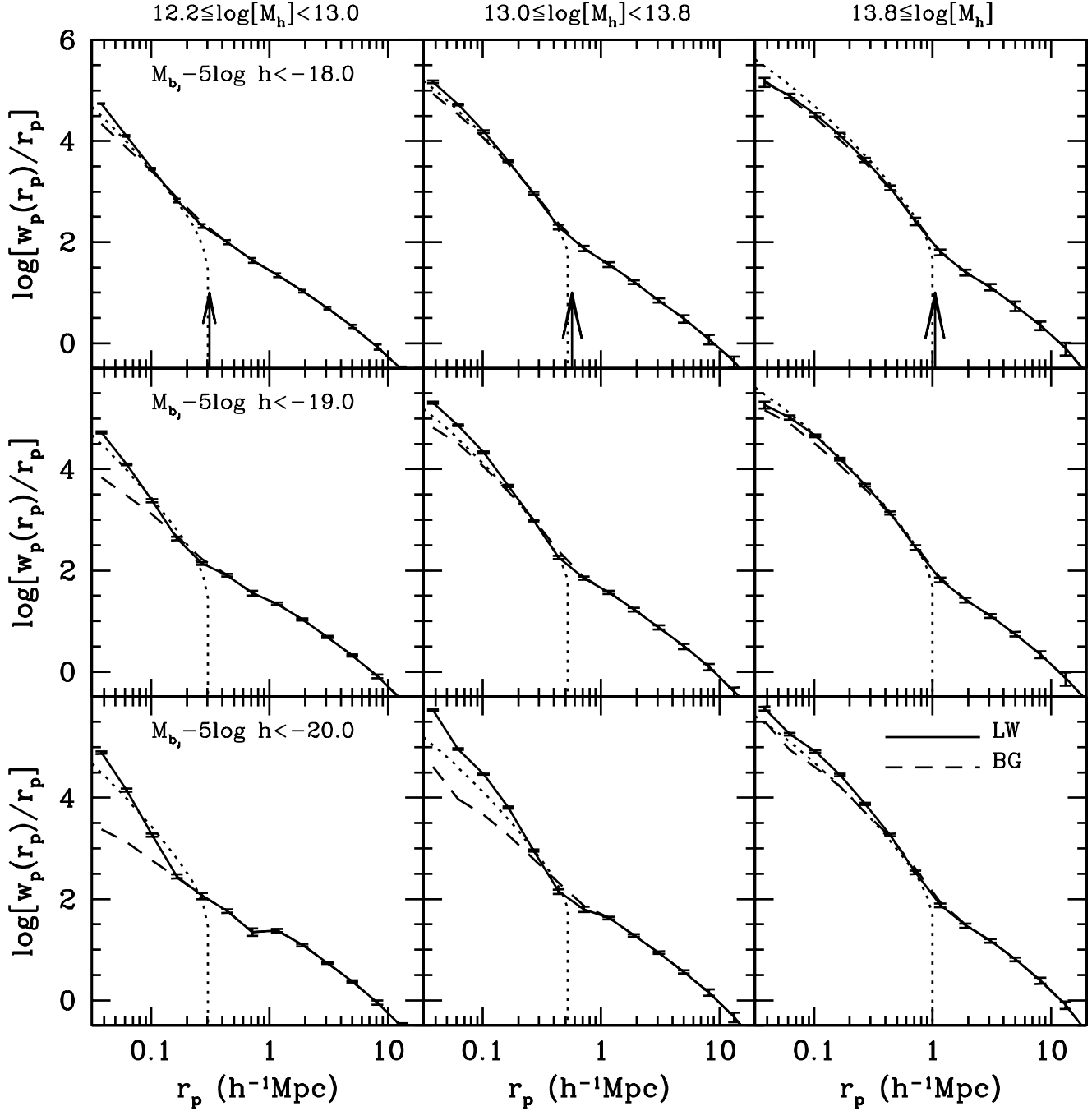,width=\hdsize}}
\caption{The projected cross-correlation function between galaxies and
  groups obtained from mock 2dFGRS catalogues.  Solid and dashed lines
  correspond  to luminosity  weighted (LW)  and brightest  galaxy (BG)
  group  centers, respectively. As  in Fig.~\ref{fig:wrp},  the dotted
  lines  illustrate the 1-halo  term corresponding  to an  NFW profile
  that belongs to a halo with a  mass that is equal to the median mass
  of the range considered.  Errorbars are obtained from the 1-$\sigma$
  variance  among  eight  independent  MGRSs,  and  thus  reflect  the
  expected variance due to cosmic scatter. }
\label{fig:mock}
\end{figure*}
\begin{figure*}
\centerline{\psfig{figure=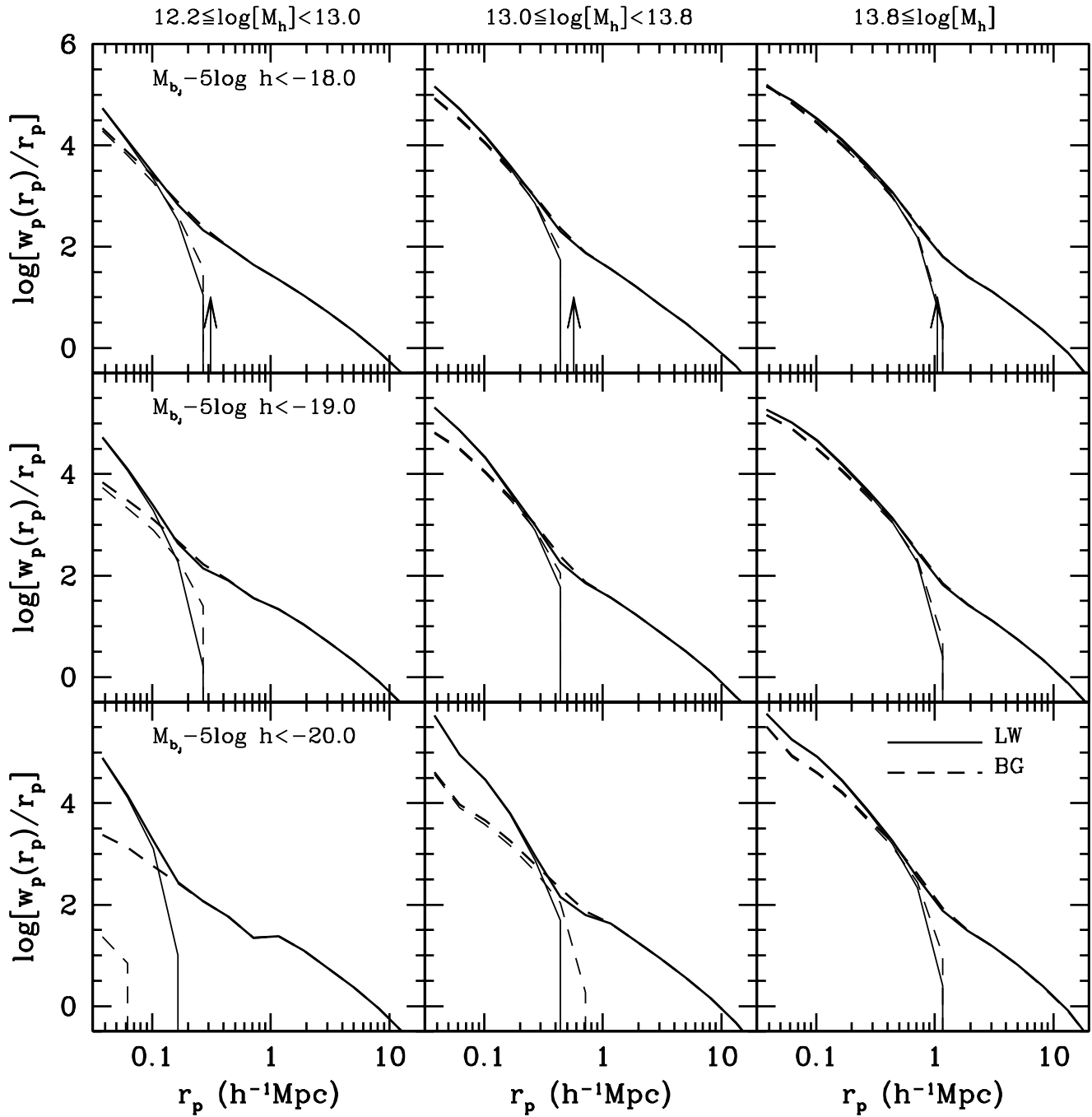,width=\hdsize}}
\caption{Same as Fig~\ref{fig:mock}, except that the `1-halo' terms are 
  plotted  separately (thin  lines). In  most cases,  the  small scale
  GHCCFs are  dominated by the  `1-halo' term and the  contribution of
  projected  `2-halo'  pairs  is  negligible.  The  exception  is  the
  `1-halo'  term of  the  GHCCF  between low  mass  haloes and  bright
  galaxies  (lower  left  panel),  which is  completely  dominated  by
  `2-halo' pairs due to projection  effects if the BG centers are used
  (see text for detailed discussion).}
\label{fig:mock_1h}
\end{figure*}
\begin{figure*}
\centerline{\psfig{figure=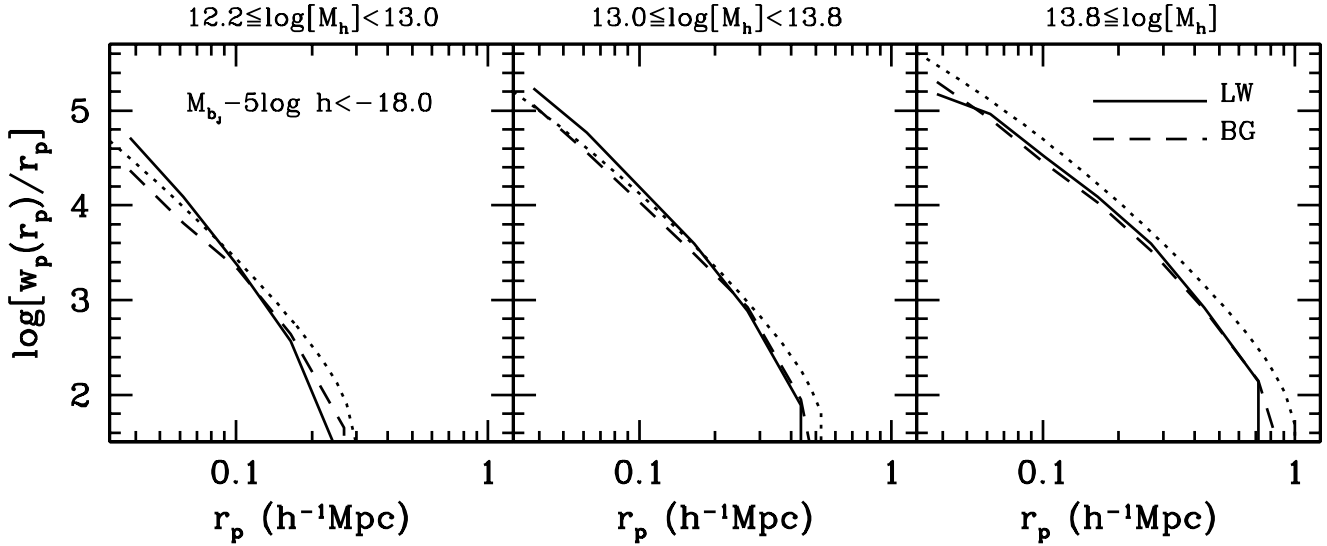,width=\hdsize}}
\caption{ The `1-halo' terms of the GHCCFs 
  obtained  from mock  catalogues  without taking  into account  fiber
  collisions and image blending.  Solid and dashed lines correspond to
  luminosity weighted  (LW) and  brightest galaxy (BG)  group centers,
  respectively.   The dotted  lines  show the  `1-halo'  terms of  the
  GHCCFs  obtained from  projecting the  NFW profiles  of  dark matter
  particles.   Note that  the shapes  of  the NFW  input profiles  are
  recovered remarkably  well by the  GHCCFs assuming BG  centers.  The
  difference in the amplitude between the NFW profile and the `1-halo'
  term of the GHCCF is due to the fact that the ratio between the mean
  number of galaxies  in a halo and the halo  mass decreases with halo
  mass for massive haloes.}
\label{fig:mock_no}
\end{figure*}
\begin{figure*}
\centerline{\psfig{figure=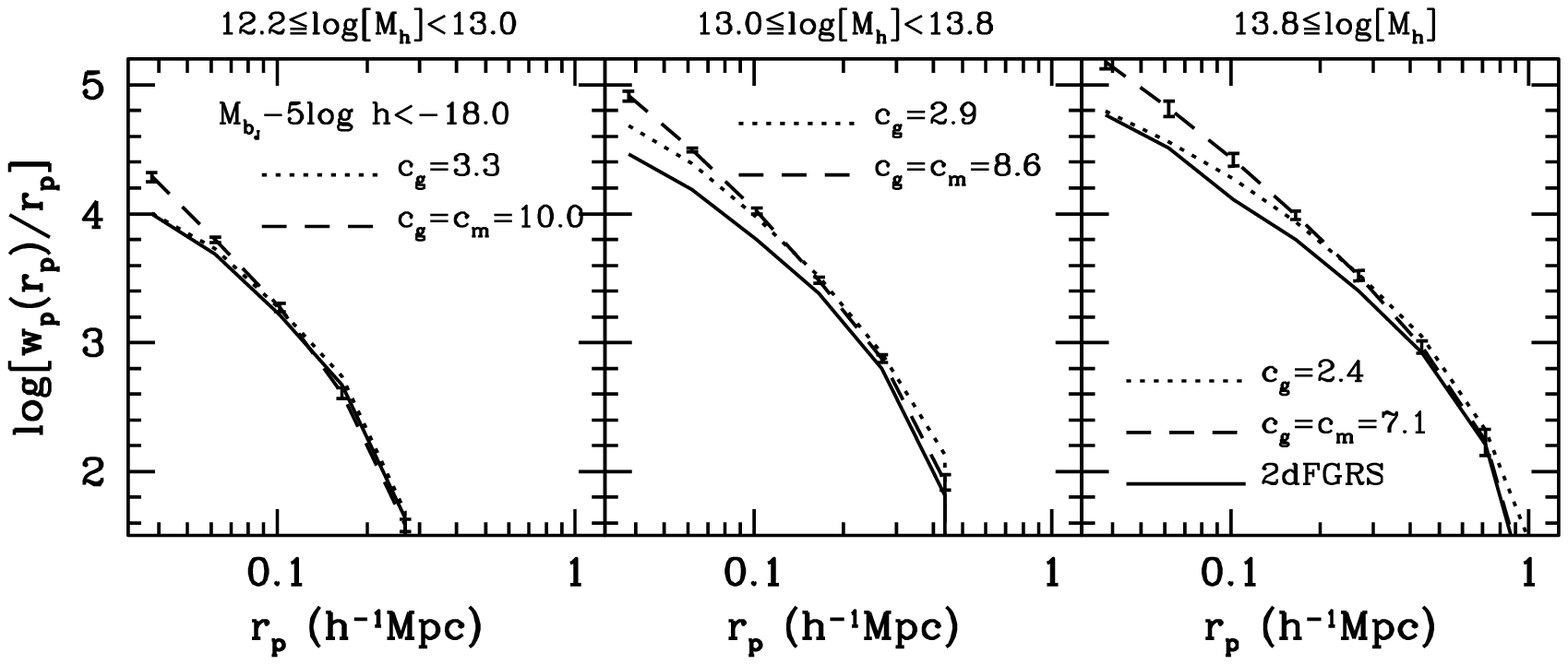,width=\hdsize}}
\caption{Comparison between the `1-halo' terms of the GHCCFs 
  obtained from the MGRSs and the 2dFGRS.  Here results are shown only
  for BG  centers.  The solid lines  are the results  from the 2dFGRS,
  while the dashed and dotted lines are the results for MGRSs assuming
  various concentrations  for the  galaxy distribution in  dark matter
  haloes, as indicated.}
\label{fig:mock_comp}
\end{figure*}

\section{Comparison with mock catalogues}
\label{sec_mock}

What can  we learn  from the above  results about the  distribution of
galaxies  in and around  dark matter haloes? Unfortunately,  a direct
interpretation is hampered by the fact that the data used suffers from
various incompleteness  effects. In particular, there  is a close-pair
incompleteness that  arises from  fiber collisions and  image blending
(Colless  \etal 2001;  Cole \etal  2001; Hawkins  \etal 2003;  van den
Bosch  \etal 2005a).  
 Obviously,  such incompleteness has important  impact on  
any pair-statistic, including the galaxy-group
cross correlation studied here, and needs to be accounted for. To this
extent we use detailed mock galaxy redshift surveys (hereafter MGRSs)
that include all these selections and incompleteness effects as present
in the real  data. The procedure of including 
fiber collision and image blending are described in 
van den Bosch \etal (2005a), and we refer the reader to 
that paper for details.   
From these MGRSs we compute the  same GHCCFs as for
the real data, allowing for a fair, one-to-one comparison.

The MGRSs  are constructed  for the 2dFGRS  by populating  dark matter
haloes  in large,  numerical  $N$-body simulations,  with galaxies  of
different luminosities. To decide what  galaxy to put in what halo, we
use the conditional luminosity function, which assures that the entire
population  of  galaxies  has  the  correct  luminosity  function  and
clustering  properties  (as  function  of  luminosity).   Within  each
individual halo,  we can modify  the spatial distribution  of galaxies
and investigate how  this impacts on the `observed'  GHCCF.  Our MGRSs
are  tailored to  resemble the  2dFGRS  as close  as possible,  taking
detailed account of the  various selection and incompleteness effects. 
A  detailed description  of  the  MGRS construction  is  given in  the
Appendix, and, in more detail, in  Yang \etal (2004) and van den Bosch
\etal (2005a).  Note  that these MGRSs have also been  used in YMBJ to
test and calibrate  the halo-based group finder used  to construct our
group catalogues.

Fig.\,\ref{fig:mock}  shows the  GHCCFs  obtained from  the MGRSs.   A
comparison with Fig.\,\ref{fig:wrp} shows  that these are very similar
to those obtained  from the 2dFGRS.  In particular,  the GHCCF between
low-mass groups and bright galaxies  is, in both cases, much shallower
than the NFW profile if the BG are used as group centers. Note that in
the MGRSs,  the brightest halo galaxy  is located at  the halo center,
while the radial number density distribution of the satellite galaxies
follows that of the dark matter particles (see Appendix).

Before we proceed to interpret  the GHCCFs obtained from the MGRSs and
compare these with those obtained  from the 2dFGRS, we address how the
GHCCF on  small scales is related  to the galaxy  distribution in dark
matter haloes.  In addition  to the observational selection bias, such
as fiber  collision and  image blending, there  are a number  of other
effects  that  may complicate  the  results  of  the observed  GHCCF.  
Firstly, the projected GHCCF may be contaminated by `2-halo' pairs due
to  projection.  Secondly,  our group  finder cannot  be  perfect, and
contaminations  can  arise  because  of interlopers.   Finally,  since
different definitions of group center  can lead to different GHCCF, it
is  necessary to  know which  definition should  be used  in  order to
extract information  about the profile of galaxy  distribution in dark
matter haloes.   With our realistic  MGRSs, we can quantify  all these
effects in detail.

In order to check the importance of projection effect, we estimate the
GHCCF functions using only `1-halo' pairs and compare the results with
the corresponding results using both `1-halo' and `2-halo' pairs.  The
results  are shown  in  Fig.\,\ref{fig:mock_1h}.  In  most cases,  the
GHCCF  on small  scale  is dominated  by  the `1-halo'  term, and  the
contamination by the projected `2-halo' pairs is negligible.  However,
if  BG centers are  used, the  GHCCF between  small haloes  and bright
galaxies is small, because most  of the satellite galaxies in low-mass
haloes are  faint.  In  this case, the  contribution by  the projected
`2-halo'  pairs   becomes  dominant  (see  the   lower-left  panel  in
Fig.\,\ref{fig:mock_1h}).

As mentioned above, our mock samples include fiber collisions and image
blending. It is interesting to see how big such effects are.
To do this, we have generated mock samples that {\it do not}
include fiber collisions and image blending, and made comparisons
between the GHCCFs obtained from such samples with
those in Fig.\,\ref{fig:mock_1h}. We found that, for small radius
where the effect is the largest, fiber collisions and image blending
reduce the GHCCF by about 10\%  if LW centers are used, while
the reduction is as large as 40\% in massive haloes if BG centers
are used. The effect is bigger for early-type galaxies in massive haloes,
presumably because these galaxies have a more concentrated distribution.

In  the MGRSs  used here,  the distribution  of satellite  galaxies in
individual haloes  is assumed to follow  the NFW profile.  In order to
examine to which  extent this input profile can  be recovered from the
GHCCF,  we  use mock  catalogues  without  taking  into account  fiber
collisions   and   image   blending.   The  results   are   shown   in
Fig.~\ref{fig:mock_no}.  As mentioned above,  if only  bright galaxies
are used, the galaxy density profile cannot be measured well for small
groups. In  order to sample the  density profile reliably,  we need to
include    faint     galaxies.    In    the     results    shown    in
Fig.~\ref{fig:mock_no},  all galaxies with  $M_{b_J}-5\log h  < -18.0$
are used.  As one can  see, the input  profile can be  reproduced. The
reproduction is better  with the BG centers, which  is consistent with
the fact that in our MGRSs  central galaxies do not sample the density
profile.

It should be pointed out that the MGRSs used above all assume that the
brightest galaxy in a halo is sitting still at the center of the halo.
As discussed in van den Bosch  et al. (2005b), such assumption may not
be  correct. In  order to  investigate the  impact of  the phase-space
distribution of the  brightest galaxies in the dark  matter haloes, we
have also  measured the  GHCCFs for the  MGRSs ${\rm  M}_{0.0}$, ${\rm
  M}_{0.5}$ and  ${\rm M}_{1.0}$  in van den  Bosch \etal  (2005b). In
${\rm M}_{0.5}$  and ${\rm M}_{1.0}$,  the brightest galaxies  are not
sitting still at  the center of the dark matter  haloes, but have both
velocity  bias  and spatial  offset.  The  velocity  bias, defined  as
$b_{vel}= \langle \sigma_{\rm cen}  \rangle / \langle \sigma_{\rm sat}
\rangle$, and spatial offset,  defined as $b_{rad}=\langle r_{\rm cen}
\rangle  / \langle  r_{\rm sat}  \rangle$,  for these  two models  are
$(b_{\rm    vel},   b_{\rm    rad})=(0.5,0.072)$    and   $(1.0,1.0)$,
respectively.  Model   ${\rm  M}_{0.0}$  has   $(b_{\rm  vel},  b_{\rm
  rad})=(0,0)$,  and is  used for  comparison.  We  found that,  if LW
centers are used, the velocity bias and spatial offset have negligible
impact on the GHCCF. Using BG centers, the results for ${\rm M}_{0.0}$
and ${\rm M}_{0.5}$ are similar,  implying that velocity bias does not
affect the  GHCCF significantly, but model ${\rm  M}_{1.0}$ predicts a
shallower  GHCCF on  small scale,  especially for  massive  haloes. In
${\rm  M}_{1.0}$,  the  brightest   galaxies  have  the  same  spatial
distribution as other galaxies, and so  the GHCCF on small scale is an
average of  the profiles around all  member galaxies. As  shown in van
den Bosch  \etal (2005b), observations  based on groups  selected from
the 2dFGRS and  the SDSS are best described  by model ${\rm M}_{0.5}$,
and ${\rm M}_{1.0}$  can be ruled out at  high confidence level. Thus,
the impact  of the phase-space distribution of  the brightest galaxies
on the GHCCF is expected to be unimportant.

With  the  above tests,  we  are  now in  a  position  to compare  our
observational results with the predictions of the MGRSs. Comparing the
GHCCFs  obtained  from  the  2dFGRS  (Fig.~\ref{fig:wrp})  with  those
obtained from  the MGRSs  (Fig.~\ref{fig:mock}), one notices  that the
former is shallower than the  latter in massive groups.  Note that the
MGRSs have  taken account of the  effects due to  fiber collisions and
image  blending. Therefore, this  discrepancy cannot  be due  to these
effects.   This   suggests  that,  in   the  $\Lambda$CDM  concordance
cosmology  considered  here,  the  distribution of  galaxies  is  less
centrally concentrated  than that of dark matter  particles.  In order
to quantify  this discrepancy, we  generate MGRSs in which  the radial
number density distribution of  satellite galaxies has a concentration
that differs from  that of their dark matter  haloes.  The results are
shown in  Fig.~\ref{fig:mock_comp}, together with the  2dFGRS results. 
It is clear that, in order to match the 2dFGRS data, the concentration
$c_{\rm g}$ of the distribution of  galaxies has to be lower than that
of the  dark matter haloes ($c_{\rm m}$)  by about a factor  of three. 
Note  that the  amplitude  of the  GHCCF  obtained from  the MGRS  for
massive haloes  is higher  than observed. This  owes to the  fact that
massive groups in the MGRSs used  are too rich (see Yang \etal 2005a). 
As discussed  in Yang  et al., this  discrepancy can be  alleviated by
either taking  $\sigma_8 \simeq 0.7$  (as opposed to $0.9$  as assumed
here), or  by considering mass-to-light ratios of  massive haloes that
are much higher than what  observations seem to suggest.  Reducing the
value of  $\sigma_8$ has the  additional advantage that is  lowers the
typical halo concentration of dark matter haloes, leading to a smaller
difference between the concentration of dark matter haloes and that of
their galaxy distribution.  However,  the change is only about $40\%$,
which is insufficient to reach the low concentrations obtained for the
galaxy  distribution.   We therefore  conclude  that  in the  standard
$\Lambda$CDM cosmology,  the distribution of galaxies  in massive dark
matter haloes must be less concentrated  than that of the dark matter. 

Using our MGRSs,  we have determined, for each of  the three ranges in
halo  mass the  concentration parameters  of the  galaxy distributions
that best match the GHCCFs of  the 2dFGRS. In addition we perform this
test  separately for  the  early- and  late-type galaxies  (classified
according  to the spectral  parameter $\eta$).   Results are  shown in
Table~\ref{tab:concen}, where we list  the resulting values of $c_{\rm
  g}$.  As  expected, the distribution of early-type  galaxies is more
centrally concentrated than that of late-type galaxies.  These results
are in qualitative  agreement with those of Collister  \& Lahav (2004,
hereafter CL04)  and Lin \etal  (2004).  Using clusters  selected from
the 2 Micron  All-Sky Survey (Jarrett 2000), Lin  \etal (2004) found a
concentration  parameter  of  $2.9\pm  0.2$ for  the  distribution  of
galaxies in  clusters, in good  agreement with our result  for massive
haloes.   Using 2dFGRS  groups selected  by Eke  et al.   (2004), CL04
obtained  the  concentration  of   the  distribution  of  galaxies  of
different types. Their  results are included in Table~\ref{tab:concen}
for  comparison.  Note  that the  CL04  results are  averages for  all
groups that contain more than  two galaxies, rather than for groups in
a given mass range. There are also other differences  
between CL04's analysis and ours. First of all, the group catalogue used by 
CL04 is different from ours. Although both are selected  from the  
2dFGRS, Yang  \etal  (2005a) have  shown that  the groups in the Eke 
\etal catalogue are systematically richer than those selected by  our 
halo-based group finder. Secondly, while our results are obtained by 
matching observations with mock catalogues that incorporate 
observational selection  effects, the results of CL04 were obtained  
by fitting the observed density  profiles directly with the NFW 
profile. However, apart from all these differences, our 
results and those of CL04 are  consistent with each other, both 
indicating that  the distribution of galaxies is  less concentrated 
than that of the dark matter.

\begin{table}
\caption{The concentration parameters of the distribution of galaxies 
($c_{\rm g}$) and dark matter particles ($c_{\rm m}$). Columns (1), 
(2), and (3) list the results for haloes
with masses  ($\log M_{\rm h}/(h^{-1}{\rm M}_{\odot})\ge 13.8$),
($13.8>\log M_{\rm h}/(h^{-1}{\rm M}_{\odot})\ge 13.0$), and 
($13.0>\log M_{\rm h}/(h^{-1}{\rm M}_{\odot})\ge 12.2$), respectively. 
For comparison, column (4) lists the results obtained by
Collister \& Lahav (2004).} \label{tab:concen}
\begin{tabular}{lcccc} 
\hline
 & $M_{\rm h,1}$ & $M_{\rm h,2}$ & $M_{\rm h,3}$ & CL04 \\
 & (1) & (2) & (3) & (4) \\
\hline\hline
All galaxies & 2.4 & 2.9 & 3.3 & $2.4 \pm 0.2$ \\
Red          & 2.5 & 3.4 & 4.0 & $3.9 \pm 0.5$ \\
Blue         & 1.1 & 1.7 & 2.2 & $1.3 \pm 0.2$ \\
\hline
Dark Matter  & 7.1 & 8.6 & 10.0& \\
\hline
\end{tabular}
\end{table}

\section{Summary}
\label{sec_summary}
 
In order to  probe the spatial distribution of  galaxies in and around
dark matter haloes, we measured the GHCCFs between galaxies and groups
(haloes) for  both the  2dFGRS and the  SDSS. The  corresponding group
catalogues are constructed using the halo-based group finder developed
in Yang \etal  (2005a).  The resulting GHCCFs show  a clear transition
from  the `1-halo' to  the `2-halo'  terms at  around the  halo virial
radius. The  `1-halo' term measures the correlation  between the group
center and the galaxies that are part of that group (i.e., it measures
the radial distribution of galaxies within their parent haloes), while
the  `2-halo'  term  measures  the  large  scale  correlation  between
galaxies that  reside in  different parent haloes.   We have  used two
different definitions  for the group  center in our estimation  of the
GHCCF; one  is the average,  luminosity weighted (LW) location  of the
member galaxies and the other  is the location of the brightest galaxy
(BG) in  the group.   The GHCCFs of  these two definitions  are almost
identical on large scales (`2-halo' term), but very different on small
scales (`1-halo' term).   The small scale GHCCF for  the BG centers is
always shallower  than that obtained using the  LW centers, especially
when  cross correlating  bright galaxies  and low  mass  haloes.  This
indicates  that  the  brightest  galaxies  in  small  haloes  play  an
important, dominant  role in the overall galaxy  distribution profile. 

We have  studied the GHCCFs  as a function  of group mass  and various
properties of the galaxies (luminosity, stellar mass, colour, spectral
type, and specific star  formation rate). Overall, more massive groups
reveal a  stronger GHCCF  than low mass  groups. On large  scales, the
GHCCF is stronger  for galaxies that are more  luminous, more massive,
red,  early-type and/or  with a  low SSFR.   All these  trends  can be
understood in terms of the mean  bias of their host haloes (i.e., more
massive  haloes are  more strongly  biased). When  using the  LW group
centers,  the GHCCFs  of these  same  galaxies are  much stronger  and
steeper than for their  counterparts, especially in massive haloes ($M
\gta 10^{13}h^{-1}M_{\odot}$).  However, when  the BG centers are used
instead,  the `1-halo'  term  of the  GHCCF  does not  show any  clear
luminosity  segregation.   This  implies  that the  strong  luminosity
segregation of galaxies observed in rich groups is almost entirely due
to their brightest central galaxy.

We compared  the GHCCFs obtained  from the 2dFGRS with  those obtained
from  detailed   mock  galaxy  redshift  surveys   (MGRSs)  that  were
constructed  to  accurately  mimic  the actual  2dFGRS.   The  overall
behavior  of the GHCCFs  obtained from  the MGRSs  is similar  to that
obtained from  the 2dFGRS,  except that the  GHCCFs of the  MGRSs have
steeper  small  scale  (`1-halo'  term) profiles  than  observed.   By
carefully  comparing  the 2dFGRS  results  with  a  set of  MGRSs,  we
determined  the  concentration  parameters  for  the  distribution  of
galaxies  (of  different types)  in  haloes  of  different masses.  In
qualitative  agreement with Collister  \& Lahav  (2004) and  Lin \etal
(2004) we find that the distribution of galaxies in dark matter haloes
is  significantly  less concentrated  than  that  of  the dark  matter
particles as predicted by the standard $\Lambda$CDM model.

                                                                               
\section*{Acknowledgment}

Numerical  simulations used  in this  paper  were carried  out at  the
Astronomical Data Analysis Center  (ADAC) of the National Astronomical
Observatory, Japan.  We  are grateful to Michael Blanton  for his help
with  the NYU-VAGC, and  thank the  2dFGRS and  SDSS teams  for making
their data publicly available. Part of the data analysis was supported  
by the Theodore Dunham, Jr. Fund for Astrophysical Research. 
                                                                               


\appendix

\section[]{Mock Galaxy Redshift Surveys}
\label{sec:AppA}

We construct MGRSs  by populating dark matter haloes  with galaxies of
different  luminosities. The  distribution  of dark  matter haloes  is
obtained from a  set of large $N$-body simulations  (dark matter only)
for  a $\Lambda$CDM  `concordance'  cosmology with  $\Omega_m =  0.3$,
$\Omega_{\Lambda}=0.7$, $h=0.7$ and  $\sigma_8=0.9$.  In this paper we
use two simulations with $N=512^3$ particles each, which are described
in more detail  in Jing \& Suto (2002).  The simulations have periodic
boundary  conditions and box  sizes of  $L_{\rm box}=100  h^{-1} \Mpc$
(hereafter  $L_{100}$) and  $L_{\rm box}=300  h^{-1}  \Mpc$ (hereafter
$L_{300}$). We  follow Yang \etal  (2004) and replicate  the $L_{300}$
box on a $4  \times 4 \times 4$ grid.  The central  $2 \times 2 \times
2$ boxes, are  replaced by a stack of $6 \times  6 \times 6$ $L_{100}$
boxes, and the  virtual observer is placed at  the center (see Fig.~11
in   Yang   \etal   2004).    This   stacking   geometry   circumvents
incompleteness problems  in the mock  survey due to  insufficient mass
resolution of  the $L_{300}$ simulations,  and allows us to  reach the
desired depth of $z_{\rm max}=0.20$ in all directions.
 
Dark  matter haloes are  identified using  the standard  FOF algorithm
with  a  linking  length   of  $0.2$  times  the  mean  inter-particle
separation.  Unbound haloes and haloes with less than 10 particles are
removed from the sample.  In Yang  \etal (2004) we have shown that the
resulting  halo mass  functions are  in excellent  agreement  with the
analytical halo mass function of Sheth, Mo \& Tormen (2001).

In order to populate the dark matter haloes with galaxies of different
luminosities,  we use the  conditional luminosity  function (hereafter
CLF), $\Phi(L \vert M)$, which gives the average number of galaxies of
luminosity $L$ that resides in a  halo of mass $M$. As demonstrated in
Yang, Mo \& van den Bosch (2003) and van den Bosch, Yang \& Mo (2003),
the CLF is  well constrained by the galaxy  luminosity function and by
the galaxy-galaxy  correlation lengths  as function of  luminosity. In
the MGRSs used here we use the CLF with ID  \# 6 given in  Table~1 of 
van den Bosch  \etal (2005a). We  have  tested that  none of  our
results depend  significantly on this  particular choice for the  CLF.

Because of the  mass resolution of the simulations  and because of the
completeness limit of the 2dFGRS, we adopt a minimum galaxy luminosity
of  $L_{\rm min} =  10^{7} h^{-2}  \Lsun$.  The  {\it mean}  number of
galaxies with $L \geq L_{\rm min}$  that resides in a halo of mass $M$
is given by
\begin{equation}
\label{totN}
\langle N \rangle_M = \int_{L_{\rm min}}^{\infty} \Phi(L \vert M) 
\, {\rm d}L
\end{equation}
In  order  to Monte-Carlo  sample  occupation  numbers for  individual
haloes, one requires the  full probability distribution $P(N \vert M)$
(with $N$ an  integer) of which $\langle N \rangle_M$  gives the mean. 
We differentiate between satellite  galaxies and central galaxies. The
total number  of galaxies per  halo is the  sum of $N_{\rm  cen}$, the
number of  central galaxies which is  either one or  zero, and $N_{\rm
  sat}$, the (unlimited) number of satellite galaxies.  We assume that
$N_{\rm sat}$ follows a  Poisson distribution and require that $N_{\rm
  sat}=0$ whenever $N_{\rm  cen}=0$.  The halo occupation distribution
is thus  specified as  follows: if $\langle  N \rangle_M \leq  1$ then
$N_{\rm sat} =  0$ and $N_{\rm cen}$ is  either zero (with probability
$P = 1 - \langle N \rangle_M$) or one (with probability $P = \langle N
\rangle_M$).  If  $\langle N \rangle_M  > 1$ then $N_{\rm  cen}=1$ and
$N_{\rm  sat}$ is drawn  from a  Poisson distribution  with a  mean of
$\langle N \rangle_M - 1$.

We follow Yang  \etal (2004) and draw the  luminosity of the brightest
galaxy in each halo from  $\Phi(L \vert M)$ using the restriction that
$L > L_1$ with $L_1$ defined by
\begin{equation}
\label{Lons}
\int_{L_1}^\infty \Phi(L\vert M) dL = 1\,.
\end{equation}
The  luminosities  of  the  satellite  galaxies are  also  drawn  from
$\Phi(L\vert M)$, but with the restriction $L_{\rm min} < L < L_1$.

The positions and velocities of  the galaxies with respect to the halo
center-of-mass are  drawn assuming that  the brightest galaxy  in each
halo resides at  rest at the center.  The  satellite galaxies follow a
number  density distribution  that is  identical to  that of  the dark
matter  particles, and  are  assumed to  be  in isotropic  equilibrium
within the dark matter potential.   To construct MGRSs we use the same
selection criteria  and observational biases as in  the 2dFGRS, making
detailed use of the survey  masks provided by the 2dFGRS team (Colless
\etal 2001; Norberg  \etal 2002).  The various steps  involved in this
process are  described in detail in  van den Bosch  \etal (2005b). The
final MGRSs  accurately match the clustering  properties, the apparent
magnitude distribution  and the  redshift distribution of  the 2dFGRS,
and  mimic all  the  various incompleteness  effects,  allowing for  a
direct, one-to-one comparison with the true 2dFGRS.

Using  a set  of  independent numerical  simulations,  we construct  8
independent  MGRSs which  we  use  to address  scatter  due to  cosmic
variance. Finally, for each MGRS  we construct group samples using the
same halo-based group finder and  the same group selection criteria as
for the 2dFGRS. These are used  to compute the GHCCFs, as described in
Section~\ref{sec_2pcf}.    The  comparison   with  the   2dFGRS  cross
correlation functions is discussed in Section~\ref{sec_mock}.

\label{lastpage}

\end{document}